\documentclass[aps,prd,twocolumn,preprintnumbers,nofootinbib,amsmath,amssymb]{revtex4-2}
\usepackage{graphicx}
\usepackage{dcolumn}
\usepackage{bm}
\usepackage{amsmath}
\usepackage{braket}
\usepackage[hidelinks]{hyperref}
\usepackage{epstopdf}
\usepackage{placeins}
\usepackage{multirow}
\usepackage{makecell}
\usepackage{cleveref}
\usepackage{xcolor}
\usepackage{dblfloatfix}
\usepackage{slashed}

\newcommand{\beq}{\begin{eqnarray}}
\newcommand{\eeq}{\end{eqnarray}}
\newcommand{\beqnn}{\begin{eqnarray*}}
\newcommand{\eeqnn}{\end{eqnarray*}}

\usepackage[font=small, labelfont=bf, textfont={small,it}]{caption}
\def\spose#1{\hbox to 0pt{#1\hss}}
\def\ltapprox{\mathrel{\spose{\lower 3pt\hbox{$\mathchar"218$}}
	\raise 2.0pt\hbox{$\mathchar"13C$}}}

\begin{document}

\title{The Roberge-Weiss transition as a probe for conformality in many-flavor QCD}

\author{Massimo D'Elia}
\email{massimo.delia@unipi.it}
\affiliation{Dipartimento di Fisica dell'Universit\`a di Pisa \& \\ INFN Sezione di Pisa, Largo Pontecorvo 3, I-56127 Pisa, Italy}

\author{Marco Nacci}
\affiliation{Dipartimento di Fisica dell'Universit\`a di Pisa \& \\ INFN Sezione di Pisa, Largo Pontecorvo 3, I-56127 Pisa, Italy}

\author{Kevin Zambello}
\email{kevin.zambello@pi.infn.it}
\affiliation{Dipartimento di Fisica dell'Universit\`a di Pisa \& \\ INFN Sezione di Pisa, Largo Pontecorvo 3, I-56127 Pisa, Italy}

\date{\today}

\begin{abstract}
  We consider the problem of identifying the onset of the conformal window for QCD with $N_f$ massless flavors in the fundamental representation, and propose a new effective method to determine it from lattice simulations. This method is based on the investigation of the so-called Roberge-Weiss transition temperature $T_{RW}$, which is encountered at specific values of the imaginary baryon chemical potential, and can also be interpreted as the inverse of the critical spatial size at which charge conjugation is spontaneously broken in a finite box. Since $T_{RW}$ corresponds to a genuine phase transition for any value of the quark masses, it is a well-defined quantity; we argue that the critical $N_f$ at which $T_{RW}$ vanishes in the chiral limit coincides with the onset of the conformal window.
  We implement our proposal by investigating QCD with $N_f = 8$ flavors, discretized via stout improved staggered fermions and the tree-level improved Symanzik pure gauge action, at Euclidean temporal extents $N_t = 8, 10, 12, 16, 24$. In this case, we find evidence that $T_{RW}$ already vanishes in the chiral limit, indicating that $N_f = 8$ is already in the conformal window.
\end{abstract}

\maketitle

\section{Introduction}

It has long been suggested that QCD with $N_f$ light flavors in the fundamental representation develops an infrared fixed point when $N_f$ exceeds a critical value $N_f^*$ but is still below the threshold at which asymptotic freedom is lost. In this regime, the theory is expected to become conformal, with the absence of dynamical mass generation, confinement, and chiral symmetry breaking \cite{Caswell:1974gg,Banks:1981nn,Miransky:1996pd,Miransky:2010kb,Appelquist:1996dq,Appelquist:1998rb,Bashir:2013zhaa}. This identifies the so-called conformal window, which for $N_c = 3$ colors corresponds to $N_f^* \leq N_f \leq 16$.

Pinning down the exact value of $N_f^*$ is important in its own right, since it is necessary for a full understanding of the phase structure of Yang–Mills theories. It is also of interest in view of the near‑conformal “walking” dynamics (i.e., slowly running coupling) expected in its vicinity, which provides an attractive framework for strongly interacting theories of physics beyond the standard model, in which the Higgs boson is interpreted as a flavor-singlet scalar, analogous to the $\sigma$ meson \cite{Holdom:1981rm,Yamawaki:1985zg,Appelquist:1986an,Appelquist:1987fc}.

Various continuum approaches provide analytical estimates of $N_f^*$ \cite{Appelquist:1996dq,Appelquist:1998rb,Ryttov:2007cx,Bashir:2013zhaa}. However, its exact value is a fully non-perturbative feature of QCD, hence a first-principles determination of $N_f^*$ can be based on numerical lattice simulations.
Indeed, several efforts have been dedicated to this issue in the last few years, exploring different strategies. These include lattice determinations, for a given value of $N_f$, of the running coupling \cite{Appelquist:2007hu, Appelquist:2009ty,Hasenfratz:2011xn,Ishikawa:2015iwa,Hasenfratz:2014rna,Fodor:2015baa,Hasenfratz:2017qyr,Hasenfratz:2019dpr,Hasenfratz:2022zsa,Witzel:2024bly}, or of the spectrum of the theory \cite{Fodor:2009wk,Fodor:2011tu,Appelquist:2011dp,Cheng:2013eu,Aoki:2013xza,LatKMI:2013bhp,LatKMI:2014xoh,LatKMI:2015ppz,LatKMI:2016xxi,LSD:2014nmn,Nagai:2015dwa,Appelquist:2016viq,LatticeStrongDynamics:2018hun}.

Another interesting approach considers instead the finite-temperature theory and the phase diagram in the $T \text{--} N_f$ plane \cite{Braun:2006jd,Miura:2012zqa,Kotov:2021hri,Cuteri:2017gci,Cuteri:2018wci,Philipsen:2019rjq,Cuteri:2021ikv,Klinger:2025xxb,Klinger:2026pbe}. It is well known that, for small values of $N_f$, the theory exhibits confinement and chiral symmetry breaking at low $T$, which then disappear at some (pseudo)-critical temperature. In the massless limit, which is the one involved in this context, the relevant symmetry is the chiral one, and a real phase transition takes place at a critical temperature $T_c$, where chiral symmetry is restored.
The interesting question is then how $T_c$ depends on $N_f$. One expects $T_c$ to be a decreasing function of $N_f$, vanishing exactly at $N_f^*$. This implies that, for $N_f \geq N_f^*$, chiral symmetry is expected to be restored already at $T = 0$. In other words, inside the conformal window no dynamically generated scale is expected to emerge, i.e., all inherent physical scales are expected to vanish, including the critical temperature.

When put into practice, this strategy encounters some challenges, the main one being that numerical simulations can only be performed at finite quark masses $m_f$, lying between the chiral limit and the heavy-mass (quenched) limit, in which dynamical quarks decouple and the theory exhibits a well-defined deconfinement transition. Therefore, from a practical point of view, to assess whether a given $N_f$ lies within the conformal window ($N_f > N_f^*$), one should determine the transition temperature as a function of $m_f$ and verify whether it vanishes in the chiral limit.

At first glance, this task seems well-defined; however, two further issues complicate its implementation:
{\em i)} at finite quark masses, QCD has no exact symmetry and therefore, in general, exhibits no genuine transition but only a crossover, with a pseudo-critical temperature whose definition may be ambiguous, depending on the chosen physical observable; 
{\em ii)} many-flavor lattice models are generally characterized by bulk transitions in the strong bare coupling regime, i.e., zero temperature transitions separating the continuum theory from unphysical phases, which may mimic an ordinary thermal chiral-symmetry-restoring transition on finite lattices, thereby requiring prohibitively expensive simulations on large lattices to disentangle the two cases.

Mainly for these reasons, the presence or absence of a thermal transition in the chiral limit of $N_f = 8$ QCD, which is believed to be very close to $N_f^*$, is still debated~\cite{Brown:1992fz,Deuzeman:2008pf,Deuzeman:2008sc,NunesdaSilva:2015jpf,Nagai:2015dwa,Aoki:2013xza,Jin:2008rc,Jin:2009mc,Jin:2010vm,LSD:2014nmn,Hasenfratz:2014rna,Schaich:2015psa,LatticeStrongDynamics:2018hun,LatKMI:2016xxi,Hasenfratz:2022zsa,Fodor:2012et,Fodor:2015baa,Fodor:2019vmw,Miura:2012zqa,Cuteri:2021ikv,Witzel:2024bly,Klinger:2025xxb,Klinger:2026pbe,DeGrand:2015zxa,Golterman:2018mfm,Golterman:2020tdq}.
\\

The purpose of the present study is to propose a novel strategy to address at least the first of the problems mentioned above. The idea is to consider a modification of the thermal theory known as the Roberge-Weiss (RW) line \cite{Roberge:1986mm}: this corresponds to a special line in the phase diagram in the presence of an imaginary baryon chemical potential~\cite{Alford:1998sd,Lombardo:1999cz,deForcrand:2002hgr,DElia:2002tig,deForcrand:2003vyj,DElia:2004ani,Giudice:2004se,Chen:2004tb,Azcoiti:2005tv}, obtained by rotating the boundary conditions of fermion fields in the Euclidean temporal direction by an appropriate global phase factor. In this setup, an exact Ising-like global symmetry exists for any value of the quark masses and is spontaneously broken at a critical temperature, known as the RW transition temperature $T_{RW}$, which has been the subject of several
  lattice~\cite{DElia:2007bkz,DElia:2009bzj,Cea:2009ba,deForcrand:2010he,Bonati:2010gi,Cea:2012ev,Wu:2013bfa,Philipsen:2014rpa,Bonati:2014kpa,Wu:2014lsa,Czaban:2015sas,Bonati:2016pwz,Bonati:2018fvg,Philipsen:2019ouy,Cardinali:2021fpu,Cuteri:2022vwk,Brandt:2022jwo,DElia:2025ybj,Zambello:2024ucs}
  and
  model~\cite{Kouno:2009bm,Sakai:2010rp,Aarts:2010ky,Rafferty:2011hd,Morita:2011eu,Kashiwa:2011td,Pagura:2011rt,Scheffler:2011te,Kashiwa:2013rm,
Nagata:2014fra,Kashiwa:2016vrl,Makiyama:2015uwa,Li:2018xgl} studies.

The RW transition is a remnant of the center-symmetry-breaking deconfinement transition that characterizes $SU(N_c)$ pure gauge theories, and available studies are consistent with chiral symmetry restoration occurring at $T_{RW}$ as well~\cite{Bonati:2018fvg,Cuteri:2022vwk}.
Our proposal is to determine $T_{RW}$ in place of the usual thermal crossover temperature, since this is a much better-defined task,
owing to the existence of exact order parameters, and then to monitor its behavior as a function of $N_f$ and of
the quark mass $m_f$.

Since numerical results show that $T_{RW} > T_c$ in all explored cases, demonstrating that $T_{RW}$ vanishes in the chiral limit for a given $N_f$ is equivalent to showing that $T_c$ vanishes as well, and hence that this value of $N_f$ already lies inside the conformal window. However, we make the stronger conjecture that $T_{RW} (m_f = 0)$, viewed as a function of $N_f$, vanishes precisely at $N_f = N_f^*$, i.e., at the onset of the conformal window.
We elaborate on this conjecture in more detail in Section~\ref{section_rw}. In brief, the rationale is twofold: on the one hand, as $T \to 0$, the boundary conditions in the temporal Euclidean direction become irrelevant; on the other hand, $N_f^*$ is characterized by the simultaneous vanishing of all dynamical scales of the theory, and hence of both $T_{RW}$ and $T_c$.
This is perhaps clearer when, as explained in Section~\ref{section_rw}, $T_{RW}$ is interpreted as the inverse of the characteristic QCD length scale below which charge conjugation is spontaneously broken in a finite box~\cite{DeGrand:2006qb,DeGrand:2007tw,Lucini:2007as,Lucini:2009kf}.
Therefore, $T_{RW}$ itself can legitimately serve as a probe of the onset of the conformal window\footnote{In principle, an exotic situation in which $T_c (N_f)$ vanishes in the chiral limit before $T_{RW} (N_f)$ cannot be ruled out. However, even in that case, the vanishing of $T_{RW}$ should still be taken as the proper signal of the onset of conformality.}.

As a first application of this novel strategy, in the following we also illustrate a preliminary analysis of $N_f = 8$ QCD discretized via stout-improved staggered fermions. Our present results, based on an analysis of the critical couplings in the bare gauge coupling -- bare quark mass plane for Euclidean temporal extents up to $N_t = 24$, are consistent with $T_{RW}$ vanishing in the chiral limit, i.e., with $N_f^* \leq 8$.

The paper is organized as follows. In Section~\ref{section_rw} we review the Roberge-Weiss transition and present our argument that it can be used as a probe of conformality. In Section~\ref{setup_sec} we provide details of the adopted lattice regularization, discuss the possible bulk transitions associated with it, and outline the employed numerical strategy. In Section~\ref{sec_results} we illustrate our results. In Section~\ref{sec:discussion} we discuss their interpretation and perform the extrapolation to zero bare quark mass in order to assess the fate of the RW transition in the chiral and continuum limits. Finally, in Section~\ref{sec_conclusions} we draw our conclusions.

\section{The RW transition and its connection to conformality}

\label{section_rw}

The Roberge-Weiss symmetry and its spontaneous breaking are closely related to the center symmetry, which characterizes $SU(N_c)$ pure gauge theories.
In the lattice theory with periodic boundary conditions in the Euclidean temporal direction, a center transformation is defined as the multiplication of all temporal links on a given time slice by the same element of the center of the gauge group. For $SU(N_c)$, the center consists of pure phase factors, $\exp( 2\pi i k / N_c)$, $k = 0, \dots, N_c -1$, corresponding to the $N_c$th roots of the identity.

This transformation is an exact symmetry of the pure gauge action, which is spontaneously broken above the deconfinement transition. A possible order parameter for this symmetry breaking is the Polyakov loop $L$, defined as the trace of the temporal Wilson line normalized by $N_c$. Indeed, $L$ is not invariant under center transformations and, above the deconfinement transition temperature $T_c$, acquires a non-zero expectation value proportional to one of the center elements.

The introduction of dynamical fermions changes this scenario. Indeed, the fermion determinant breaks center symmetry explicitly, since it contains a direct coupling to the Polyakov loop (as can be easily proven from the loop expansion of the determinant). For standard thermal boundary conditions, this coupling favors a Polyakov loop aligned along the positive real axis in the complex plane, much like an external magnetic field in a spin model.

However, the situation changes when the fermionic temporal boundary conditions are twisted by a phase factor $\exp(i \theta_q)$.
This occurs, for instance, in the presence of a purely imaginary baryon chemical potential~$\mu_B$~\cite{Alford:1998sd,Lombardo:1999cz,deForcrand:2002hgr,DElia:2002tig,deForcrand:2003vyj,DElia:2004ani,Giudice:2004se,Chen:2004tb,Azcoiti:2005tv},
for which $\theta_q = {\rm Im} (\mu_B) / (N_c T)$ for all quark flavors.
Indeed, in this case, the Polyakov loop enters the fermion determinant multiplied by $\exp(i \theta_q)$, so it tends to be aligned along a direction forming an angle $-\theta_q$ with the real axis. If this angle corresponds exactly to the boundary between two center sectors, i.e. for $\theta_q = (2 k + 1) \pi / N_c$ with $k \in \mathbb{Z}$ (which defines the so-called RW line), then the system exhibits a $Z_2$ global symmetry, a remnant of the full center group, which can be spontaneously broken at high $T$: this is the RW transition. 
For odd values of $N_c$, one of the Roberge-Weiss lines coincides with $\theta_q = \pi$ and may be interpreted as a finite-size transition (due to the switch from antiperiodic to periodic boundary conditions for fermions) accompanied by the spontaneous breaking of charge conjugation~\cite{DeGrand:2006qb,DeGrand:2007tw,Lucini:2007as,Lucini:2009kf}. In this case, suitable order parameters are the imaginary part of the Polyakov loop, as well as the imaginary part of the baryon number.

The RW transition and the critical behavior around it have been widely studied in various lattice
  investigations~\cite{DElia:2007bkz,DElia:2009bzj,Cea:2009ba,deForcrand:2010he,Bonati:2010gi,Cea:2012ev,Wu:2013bfa,Philipsen:2014rpa,Bonati:2014kpa,Wu:2014lsa,Czaban:2015sas,Bonati:2016pwz,Bonati:2018fvg,Philipsen:2019ouy,Cardinali:2021fpu,Cuteri:2022vwk,Brandt:2022jwo,DElia:2025ybj,Zambello:2024ucs},
  mostly for 2 or 3 flavors. It has generally been found that $T_{RW}$ lies above the pseudo-critical crossover temperature
of QCD; for instance, $T_{RW} = 208(5)$~MeV for $N_f = 2+1$ QCD with physical quark masses. Depending on the flavor spectrum, the transition can be second order in the $3D$ Ising universality class, first order, or tricritical.
Moreover, current evidence is consistent with the fact that, in the massless case, chiral symmetry restoration coincides with $T_{RW}$~\cite{Bonati:2018fvg,Cuteri:2022vwk}.
\\

As already explained in the Introduction, our program is to investigate the behavior of $T_{RW}$ as a function of the number of flavors and their mass spectrum, in order to pinpoint the critical value $N_f^{RW}$ for which $T_{RW}$ vanishes in the chiral limit. The clear advantage over the usual approach, based on the pseudo-critical temperature $T_c$, is that in this case one deals with a true phase transition for any value of the quark masses, whose location is unambiguous and can, in principle, be determined with arbitrary precision.
Thus, assuming that the relation $T_{RW} > T_c$ is always valid\footnote{We emphasize that this hypothesis is not only based on previous studies in the literature, but is also supported by our current numerical results, which show that, for finite quark masses, the transition shifts towards higher critical couplings, corresponding to higher temperatures, as one moves from zero chemical potential to the RW line;
see the discussion in Section~\ref{sec:results_A}.}, one can conclude that, if $T_{RW} = 0$ in the chiral limit, then $T_c$ vanishes as well, i.e., $N_f^{RW} \geq N_f^*$.

\begin{figure}[t!]
  \centering
  \includegraphics[width=0.9\linewidth, clip]{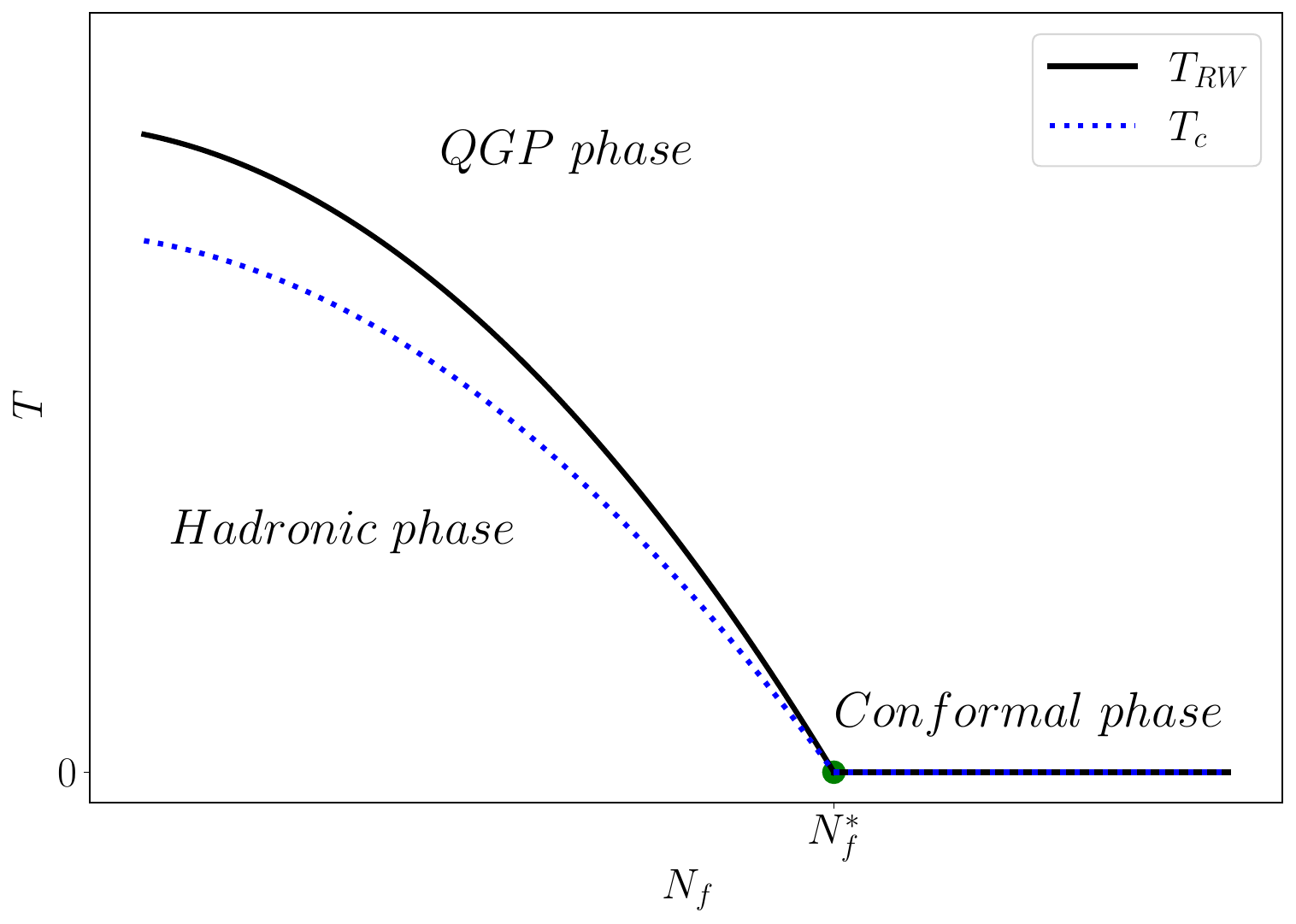}
  \caption{Conjectured behavior of $T_c$ and $T_{RW}$ in the chiral limit
    as a function of $N_f$. $T_{RW}$ is always larger than $T_c$,
      consistently with what has been observed by numerical simulations, however the two quantities
      vanish at the same point, consistently with the disappearance of all characteristic lengths in the conformal
    window.}
  \label{fig:diagramma_t_nf}
\end{figure}

However, as mentioned above, we can make a stronger claim. In the conformal window, all inherent physical scales are expected to disappear, so $T_{RW}$ must vanish as well.
This suggests that one actually has $N_f^{RW} = N_f^*$, i.e., that $T_c$ and $T_{RW}$ vanish simultaneously, possibly while keeping a fixed ratio between them. Another argument in support of this claim is that the standard thermal theory and the one along the RW line differ only by a boundary condition along the Euclidean temporal direction: in the zero temperature limit, where the temporal extent goes to infinity, that boundary condition must become irrelevant.

Our conjecture is better illustrated in Fig.~\ref{fig:diagramma_t_nf}, where both $T_c$ and $T_{RW}$ are shown in the chiral limit in a tentative phase diagram in the $T \text{--} N_f$ plane. According to this conjecture, $T_{RW}$ serves as a direct probe of the onset of the conformal window.
In what follows, we will start our program by considering the $N_f = 8$ case. We will work at $\theta_q = \pi$, so that the imaginary part of the Polyakov loop can be used as the order parameter. More generally, for the purposes of our program, we will not be interested in the order of the transition, but only in its location.

\section{Numerical set-up}

\label{setup_sec}

We study the lattice formulation of eight-flavor QCD, using the tree-level improved Symanzik pure gauge action~\cite{weisz,curci} and the
stout improved rooted staggered fermion action~\cite{kogut-susskind,morningstar}. The partition function is
\begin{equation}
  Z=\int{[DU]}\,e^{-S_{YM}} \det{(M_{st}^8)}^\frac{1}{4},
\end{equation}
where $[DU]$ is the Haar measure over the gauge links,  while the staggered fermion matrix and the improved Symanzik gauge action
are respectively
\beq
{(M_{st\ })}_{ij} & =&  \hat{m}\delta_{ij}+\sum_{\nu=1}^{4}\frac{\eta_{i;\nu}}{2}
\left( e^{a \mu \delta_{\nu,4}} U^{(2)}_{i;\nu}\delta_{i, \, j-\hat{\nu}} 
 - \right. \nonumber \\
 & &  \left. e^{-a\mu \delta_{\nu,4}} U^{(2)\dagger}_{i-\hat{\nu};\nu}\delta_{i, \, j+\hat{\nu}}\right) \nonumber \\
\label{dirac_operator}
  S_{YM} &=& -\frac{\beta}{3} \, \sum_{\substack{i \\ \mu\ne\nu}}\left(\frac{5}{6}W^{1\times1}_{i,\mu\nu}-\frac{1}{12}W^{1\times2}_{i,\mu\nu}\right) \mbox{.}
\eeq
The indices $i,j$ and $\mu,\nu$ denote, respectively, the lattice sites and directions of the gauge links, while $\eta_{i;\nu}$ are the staggered quark phases,  $U^{(2)}_{i;\nu}$ are the twice stout smeared links (with isotropic smearing parameter $\rho=0.15$), and $W^{1\times 1,2}_{i,\mu\nu}$ are the real parts of the trace of the link products along the $1\times1$ and $1\times2$ rectangular closed loops. The dimensionless constants $\beta$ and $\hat m = a m$ are the inverse gauge coupling and the bare quark mass in lattice units, respectively. The chemical potential is denoted by $\mu$ and in the following we will make use of the standard notation $\hat{\mu} = \mu / T$.

In addition to the phases associated with the RW symmetry, periodic/antiperiodic boundary conditions are imposed in the Euclidean temporal direction for bosonic/fermionic fields. 
{Gauge configurations are generated using the RHMC algorithm~\cite{Kennedy:1998cu,Clark:2006fx} with the GPU code \texttt{OpenSTaPLE}~\cite{openstaple,Bonati:2017ovw,Bonati:2018wqj}, while the \texttt{nissa}~\cite{nissa} code is used to integrate the Wilson flow.}

\subsection{Order parameter for the single-site shift symmetry breaking}

The bulk transitions observed {in many-flavor lattice QCD simulations using staggered fermions} are understood as transitions to an unphysical $\slashed{S}_4$ phase characterized by the spontaneous breaking of the single-site shift symmetry $S_4$ of the staggered action \cite{Cheng:2011ic,Deuzeman:2012ee}. The symmetry acts as
\beq
U_{n;\mu} \to U_{n+\mu;\mu} \\ \nonumber
\chi_n \to \eta_{n;\mu} \chi_{n+\mu}  \\ \nonumber
\bar{\chi}_n \to \eta_{n;\mu} \bar{\chi}_{n+\mu}
\eeq
where $U_{n;\mu}$, $\eta_{n;\mu}$ and $\chi_n$ are the gauge links, the staggered phases and the staggered fields, respectively.
The $\slashed{S}_4$ phase is a lattice-artifact phase that appears at strong coupling in theories with many flavors of staggered fermions and is not associated with any continuum symmetry of QCD.

To monitor the breaking of the $S_4$ symmetry, we employ the order parameter introduced in~\cite{Cheng:2011ic},
\begin{equation}
P_\mu = \langle ReTr\square_{n,\mu} -  ReTr\square_{n+\mu, \mu} \rangle_{n_\mu~even}.
\end{equation}
Here, $\square_{n,\mu}$ denotes the plaquette based at the lattice site $n$ with a direction along $\mu$, and the expectation value is taken over all plaquettes living on lattice sites whose $\mu$ component is even. In other words, this quantity is the expectation value of the difference between plaquettes originating at neighboring lattice sites displaced by one lattice unit along the $\mu$ direction.

When the single‑site shift symmetry is unbroken, this difference vanishes. A non-zero value of $P_\mu$ in one or more directions signals that the symmetry has been spontaneously broken and identifies the exotic $\slashed{S}_4$ phase. Since $P_\mu$ may acquire a non-zero expectation value only in certain directions and these directions may occasionally change, it is more convenient to monitor the combined parameter $\sqrt{P_\mu P_\mu}$.

\begin{figure}[t!]
  \centering
  \includegraphics[width=0.9\linewidth, clip]{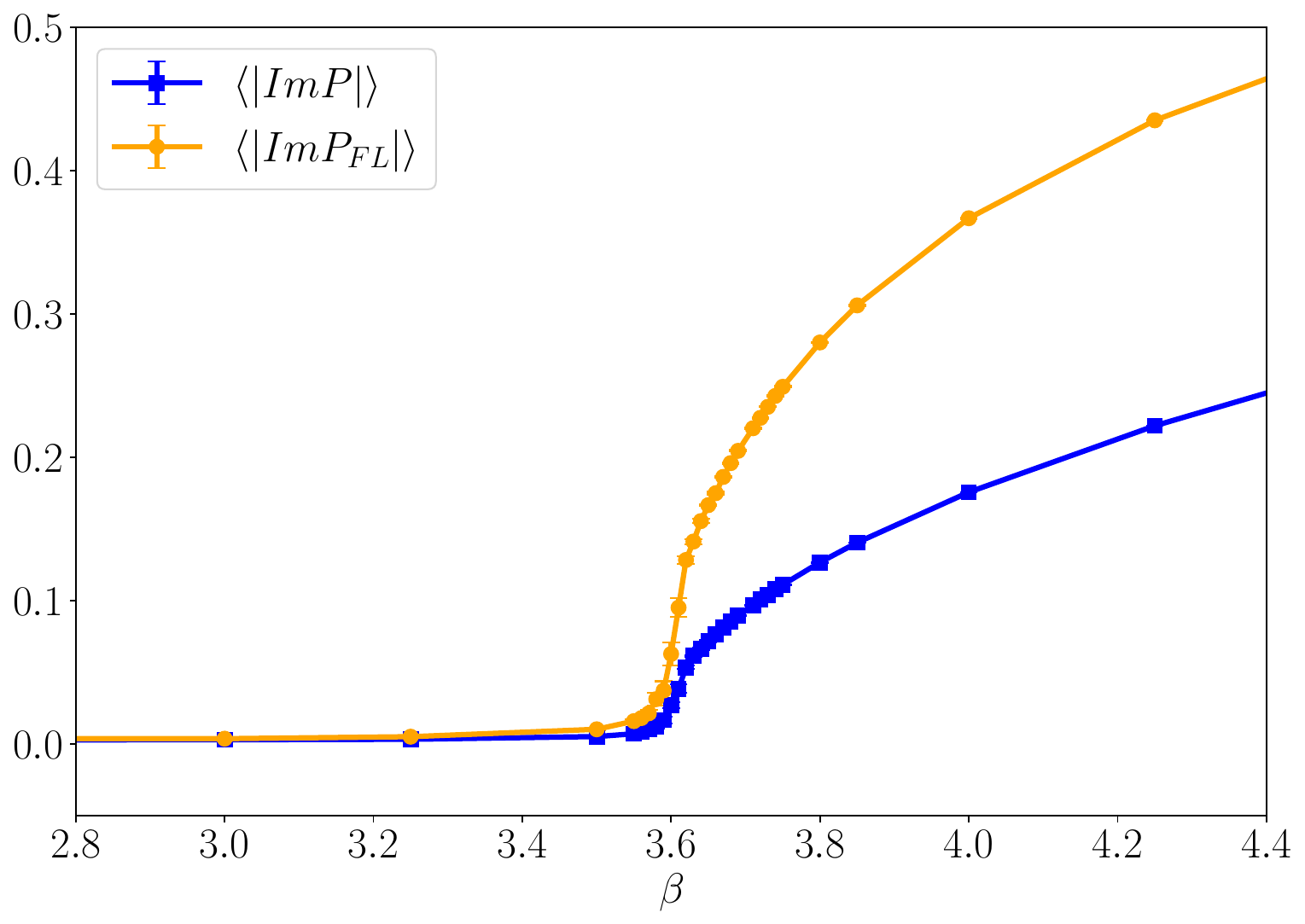}
  \includegraphics[width=0.9\linewidth, clip]{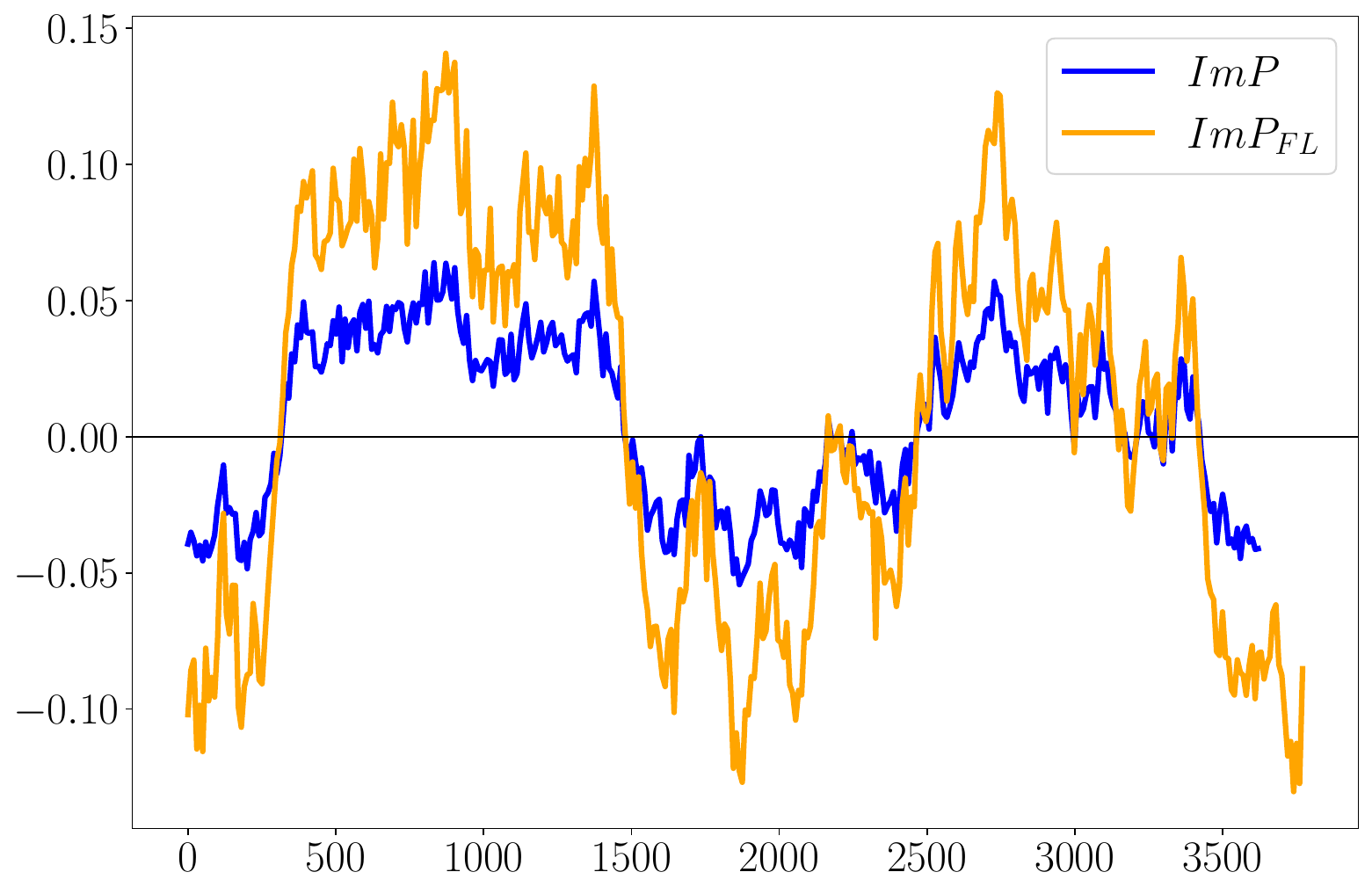}
  \caption{Cross-checks in $N_f=2$ QCD: Polyakov loop and the Wilson-flowed Polyakov loop as a function of $\beta$ (top); Monte Carlo history of the Polyakov loop and the Wilson-flowed Polyakov loop at $\beta = 3.6$ (bottom). Parameters used: $16^3 \times 4$ lattice, $\hat{m}=0.01$, $\hat{\mu} = i\pi$.}
  \label{fig:flowed_polyakov_crosscheck}
\end{figure}

\begin{figure}[t!]
  \centering
  \includegraphics[width=0.9\linewidth, clip]{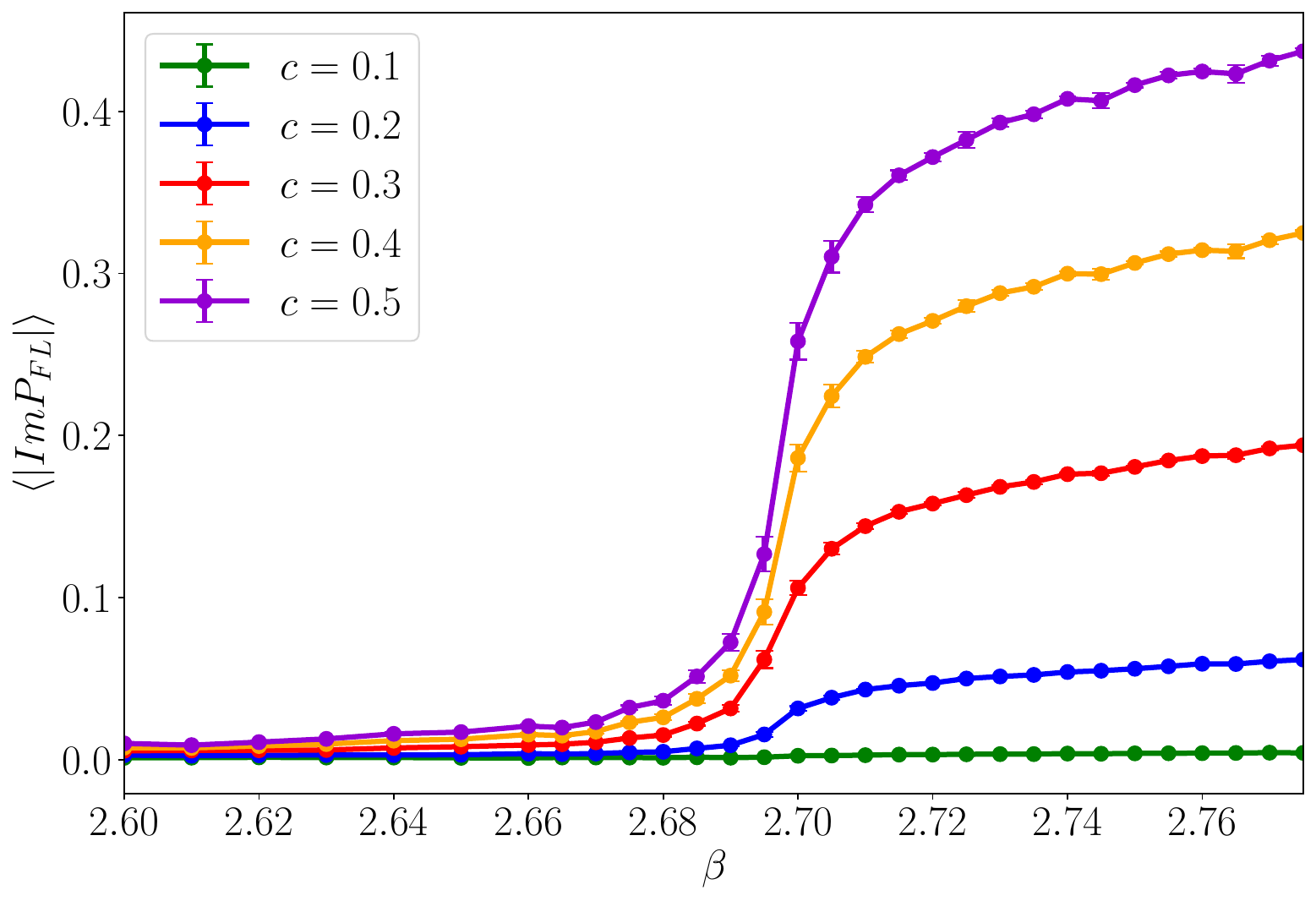}
  \includegraphics[width=0.9\linewidth, clip]{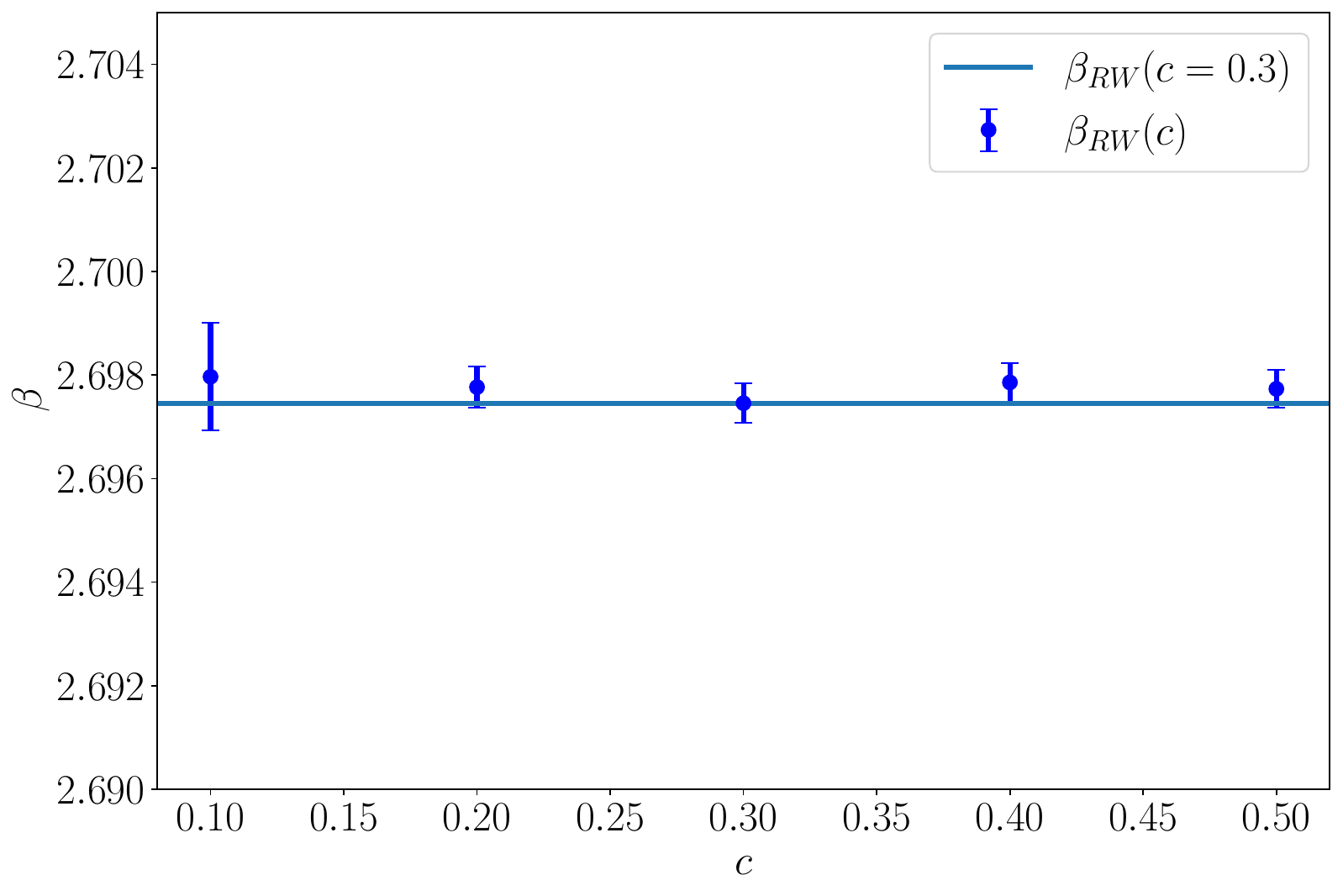}
  \caption{Cross-checks in $N_f=8$ QCD: Wilson-flowed Polyakov loop as a function of $\beta$ for different values of $c$ (top); critical values of $\beta$ estimated for different values of $c$ (bottom). Parameters used: $28^3 \times 10$ lattice, $\hat{m}=0.0125$, $\hat{\mu}=i\pi$.}
  \label{fig:c_dependence}
\end{figure}

\subsection{Improved order parameter for the Roberge-Weiss transition}

The natural order parameter for the Roberge-Weiss transition is the imaginary part of the Polyakov loop.
However, the Polyakov loop suffers from a deteriorating signal-to-noise ratio as the temporal extent $N_t$ increases. In this work we consider lattices up to $N_t=24$ and we observe that this issue becomes significant for $N_t > 8$.

Previous studies~\cite{Schaich:2015psa,LatticeStrongDynamics:2018hun}, investigating the deconfinement transition at zero chemical potential, have shown that applying the Wilson flow~\cite{Luscher:2013vga} substantially improves the quality of the Polyakov loop signal by smoothing ultraviolet fluctuations. Here we adopt the same strategy for the Roberge-Weiss transition. As a preliminary step, we perform some cross-checks to assess the reliability of the flowed observable for studying phase transitions at imaginary chemical potentials.

We begin with a controlled case, namely $N_f=2$ QCD, which is theoretically better understood than $N_f=8$ QCD. Following the prescription of Refs.~\cite{Schaich:2015psa,LatticeStrongDynamics:2018hun}, the flow time $t$ has been fixed through the dimensionless parameter $c=\sqrt{8t}/N_t=0.3$. The results are illustrated in 
Fig.~\ref{fig:flowed_polyakov_crosscheck}.
The top panel displays the observables as a function of $\beta$, while the bottom panel shows their Monte Carlo histories close to the critical coupling.
Comparing the unimproved Polyakov loop (blue) with the Wilson‑flowed Polyakov loop (orange), we observe a clear improvement in the signal quality of the flowed observable. Both observables locate the RW transition at the same critical coupling, $\beta_{RW}=3.6092(15)$.

We then extend the analysis to $N_f=8$ QCD, comparing the Wilson-flowed Polyakov loops at different flow times. As illustrated in the top panel of Fig~\ref{fig:c_dependence}, the Wilson flow again improves the signal quality. The bottom panel shows how the critical coupling exhibits negligible dependence on $c$; in particular $\beta_{RW} (c)$ is fully compatible with $\beta_{RW} (c = 0.3) = 2.6975(04)$ for all explored values of $c$.

We stress that in these cases, in contrast to the determinations of $\beta_{RW}$ discussed below, the reported errors are purely statistical, i.e., they do not account for systematics such as those associated with the choice of the fitting function, since we expect that, for each chosen fitting function, determinations obtained at different flow times should be consistent with each other.

\FloatBarrier

\section{Numerical Results}

\label{sec_results}

We study the phase diagram of eight-flavor QCD by performing numerical simulations on lattices with temporal extents $N_t=8, 10, 12, 16, 24$. The $N_s/N_t$ ratio is kept between $2$ and $3$, the bare quark mass $\hat m$ is varied in the range $0.0025$--$0.08$, while the quark chemical potential is set to $\hat{\mu} = i \theta_q = i \pi$. We simultaneously measure two observables: the imaginary part of the Polyakov loop $|\textrm{Im} P|$, which is used to identify the Roberge-Weiss phase transition, and the order parameter $\sqrt{P_\mu P_\mu}$, which is used to determine the boundaries of the exotic $\slashed{S}_4$ phase. 

For $N_t = 8$, we perform additional simulations at zero chemical potential to evaluate the effect of adding an imaginary chemical potential on the phase transitions. In this case, we measure the chiral condensate $\braket{\bar{\psi}\psi}$, the real part of the Polyakov loop $\textrm{Re} P$ and the exotic order parameter $\sqrt{P_\mu P_\mu}$ to monitor the chiral, deconfinement and exotic transition, respectively.
{We use $4$ random sources to measure the chiral condensate through noisy estimators \cite{Dong:1993pk}.} For $N_t > 8$ the unimproved Polyakov loop is too noisy to reliably locate any phase transition; we therefore employ the Wilson-flowed Polyakov loop, as discussed in the previous section, with the flow time fixed by the prescription $c=0.3$.
The parameters used in the simulations are summarized in Tab.~\ref{tab:parameters}.

\begin{table}[b!] \centering \begin{tabular}{|c|c|c|} \hline $N_t$ & $N_s$ & $\hat{m}$ \\ \hline 8 & 24 & 0.0100, 0.0125, 0.013844, 0.02, 0.03, 0.04, 0.06, 0.08 \\ \hline 10 & 20, 28 & 0.01, 0.0125, 0.02, 0.04 \\ \hline 12 & 36 & 0.0100, 0.0125, 0.0200, 0.0400 \\ \hline 16 & 32 & 0.0050, 0.0100, 0.0125, 0.0200, 0.0400 \\ \hline 24 & 48 & 0.0025, 0.0050, 0.0100, 0.0125, 0.0200 \\ \hline \end{tabular} \caption{Lattice extents and bare quark masses used in this work.} \label{tab:parameters} \end{table}

\begin{figure}[t!]
  \centering
  \includegraphics[width=0.9\linewidth, clip]{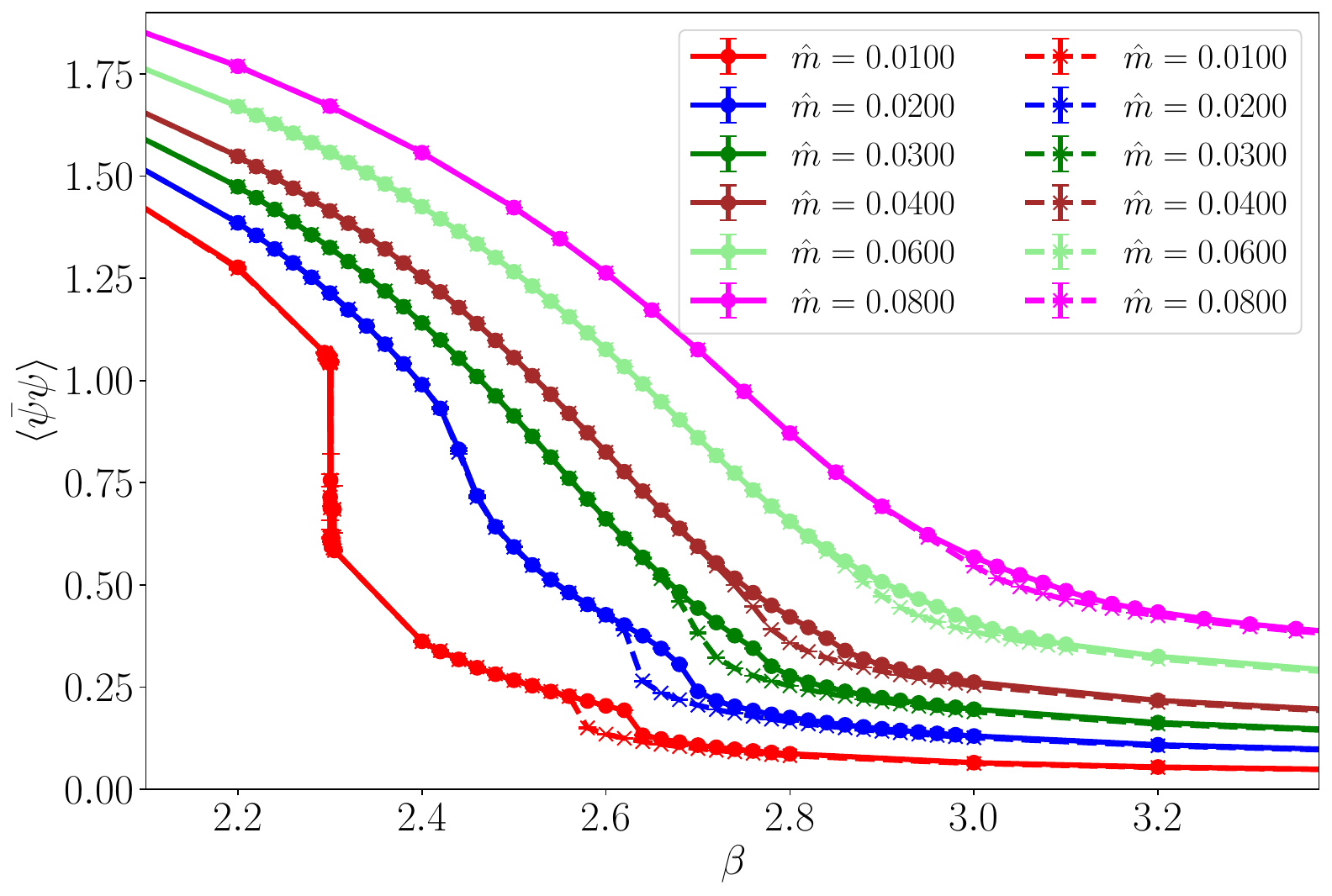}
  \caption{Chiral condensate for different masses and $\hat{\mu} = 0$ (dashed), $i\pi$ (solid), from simulations on a  $24^3 \times 8$ lattice.}
  \label{fig:ffmu0pi_menomasse}
\end{figure}

\subsection{Results for $N_t = 8$}

\label{sec:results_A}

For $N_t=8$ we perform numerical simulations at both $\hat{\mu} = 0$ and $\hat{\mu} = i \pi$, for several values of the quark mass, measuring the Polyakov loop, the chiral condensate and the order parameter associated with the exotic phase. 

The results in Fig.~\ref{fig:ffmu0pi_menomasse} compare the chiral condensate at $\hat{\mu} = 0$ (dashed lines) and $\hat{\mu} = i \pi$ (solid lines) for several quark masses ranging from $\hat m=0.01$ 
to $\hat m=0.08$.
For small quark masses, the chiral condensate exhibits the characteristic two-jump structure observed in previous studies \cite{Cheng:2011ic,Deuzeman:2012ee,Fodor:2012et,NunesdaSilva:2015jpf}. As the mass increases, the transitions become smoother and the second jump becomes increasingly difficult to resolve. The introduction of an imaginary chemical potential leaves the first jump unaffected but shifts the second jump to weaker coupling.

\begin{figure}[t!]
  \centering
  \includegraphics[width=0.9\linewidth, clip]{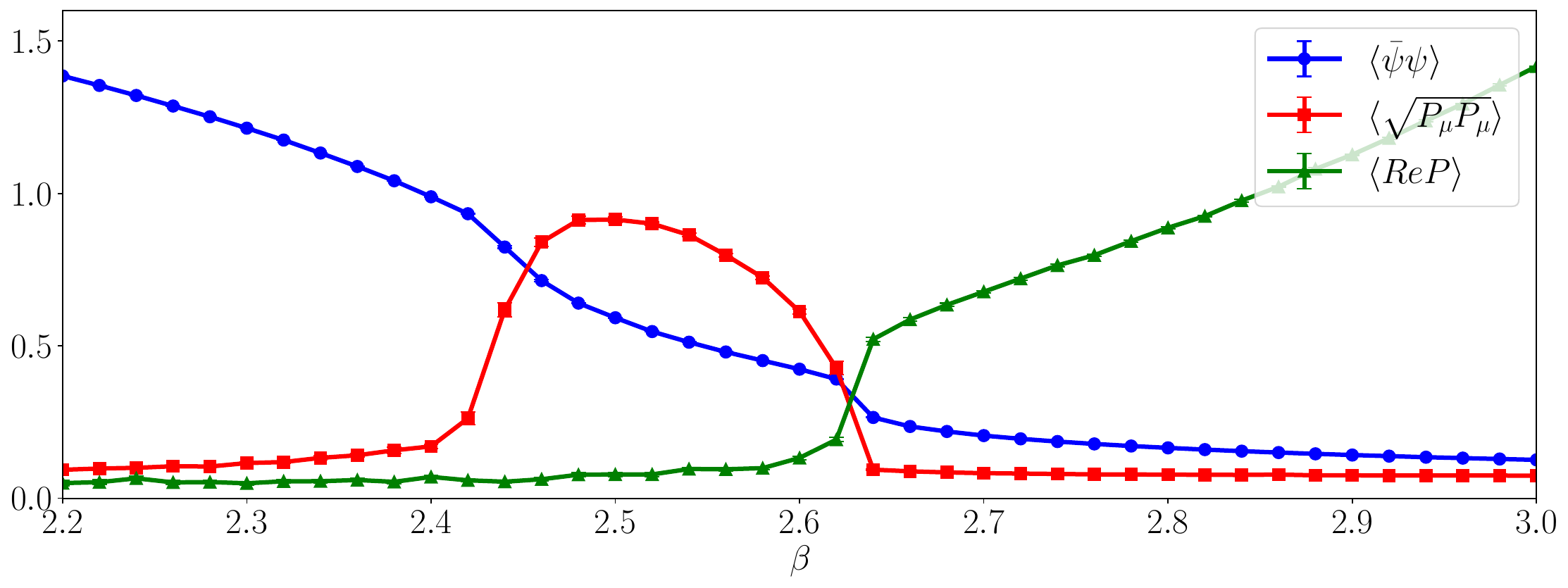}
  \includegraphics[width=0.9\linewidth, clip]{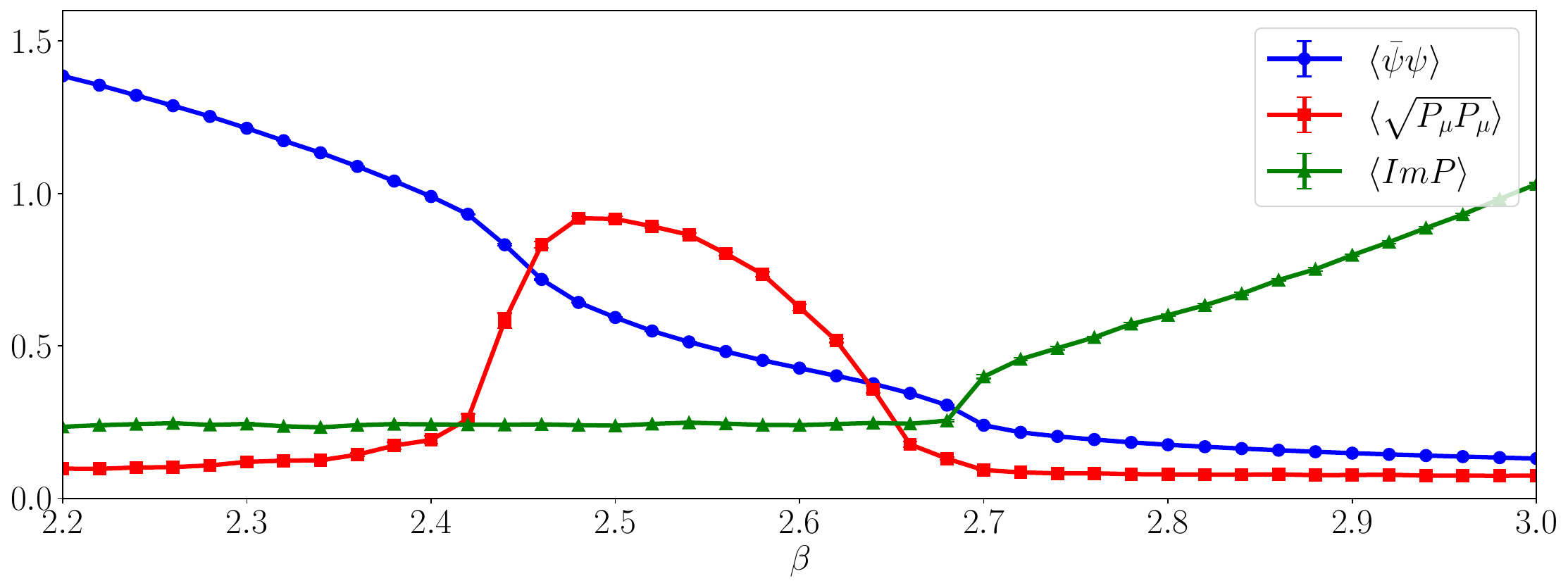}
  \caption{Chiral condensate compared with other observables for $\hat m = 0.02$ and $\hat{\mu} = 0$ (top), $i\pi$ (bottom), from simulations on a $24^3 \times 8$ lattice.}
  \label{fig:ffmu0piL24T8}
\end{figure}

To better understand the origin of this structure,  Fig.~\ref{fig:ffmu0piL24T8} shows the three observables for a representative mass, $\hat m = 0.02$. The top panel shows the results at zero chemical potential. The first jump in the chiral condensate (blue) coincides with the strong-coupling-to-exotic bulk transition (signaled by the exotic order parameter, red), while the second jump coincides with the exotic-to-weak-coupling bulk transition and with the rise of the real part of the Polyakov loop (green). Thus, the chiral condensate is sensitive to both the first bulk transition and the deconfinement transition, which in this case coincides with the second bulk transition. 

The bottom panel of Fig.~\ref{fig:ffmu0piL24T8} shows instead the corresponding results at $\hat{\mu} = i \pi$. Introducing an imaginary chemical potential shifts the critical coupling associated with the Polyakov loop towards weaker couplings, corresponding to higher temperatures. This observation aligns with the expectation that the Roberge-Weiss transition occurs at a higher temperature than the transition at zero $\hat{\mu}$. The chiral condensate still shows sensitivity to both the first bulk transition and the transition signaled by the Polyakov loop. At this particular mass, the Roberge-Weiss phase transition is distinct from the exotic-to-weak-coupling bulk transition; however, as we will see shortly, they will eventually merge at smaller masses.

\begin{figure}[t!]
  \centering
  \includegraphics[width=0.9\linewidth, clip]{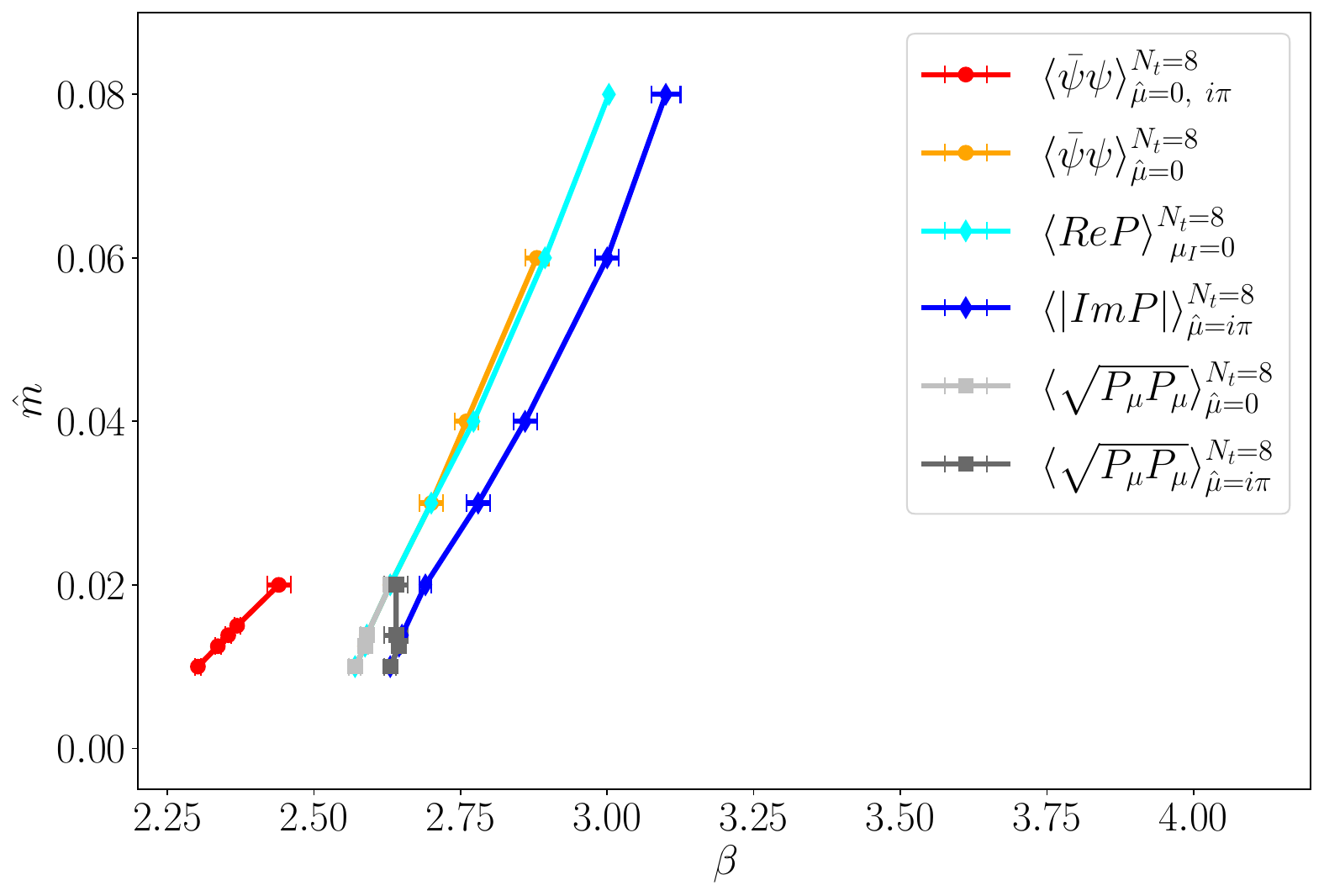}
  \caption{Phase diagram of $N_f=8$ QCD on $N_t=8$ lattices at $\hat{\mu} = 0$ and $\hat{\mu}= i \pi$. Phase transitions determined from different observables are shown with different colors. In some cases, the phase transitions fall on top of each other, so that the different symbols are hardly visible.}
  \label{fig:cc_l_vs_pp_nt8}
\end{figure}

Fig.~\ref{fig:cc_l_vs_pp_nt8} summarizes these observations and presents the $(\beta, \hat m)$ phase diagram for $N_t=8$.
{Each data point represents the critical coupling determined from the corresponding observable. We identify the critical coupling as the inflection point of an arctangent-like fit to the data. When the data lacks sufficient resolution for a reliable fit, we estimate the critical point and its uncertainty by calculating the average and half-difference of the two points between which the concavity changes.} 

{At zero chemical potential, the transition signaled by the second jump in the chiral condensate (orange) coincides with the deconfinement transition signaled by the real part of the Polyakov loop (cyan). For $\hat m\leq 0.02$ it also coincides with the right boundary of the exotic phase as determined by $\sqrt{P_\mu P_\mu}$ (gray). We determine this boundary only for $\hat m \leq 0.02$, since the signal for $\sqrt{P_\mu P_\mu}$ fades at larger masses, making the boundary difficult to resolve. Moreover, its precise location at larger masses is not essential for the present work, which focuses on the low-mass region.}
{The first jump in the chiral condensate (red) coincides with the left boundary of the exotic phase at $\hat{m}=0.02$ (see Fig.~\ref{fig:ffmu0piL24T8}), and most likely for $\hat{m}<0.02$ as well\footnote{{We have not thoroughly determined the left boundary of the exotic phase by measuring the $\sqrt{P_\mu P_\mu}$ order parameter, since the region to the left of the exotic phase is disconnected from continuum QCD and is therefore not relevant to the present work. However, in some runs preliminary to this work, we have observed that the first jump in the chiral condensate is insensitive to $N_t$, suggesting that it is associated with a bulk transition.
}}}.

Upon introducing the imaginary chemical potential, {the transition at stronger coupling signaled by the first jump of the chiral condensate remains unaffected (see also Fig.~\ref{fig:ffmu0pi_menomasse}).} However, both the {exotic-to-weak-coupling} bulk transition (black) and the transition signaled by the imaginary part of the Polyakov loop (blue){, which we identify as the Roberge-Weiss transition,} shift toward weaker couplings. These transitions are distinct at larger masses, but merge around $\hat m \sim 0.01$.
Since for $\hat m < 0.01$ the Roberge-Weiss transition coincides with the bulk transition and the latter is not connected to continuum QCD, it is unclear whether the Roberge-Weiss transition survives in the chiral limit. In the following, we address this issue by extending our study to larger temporal extents $N_t$.

\FloatBarrier

\subsection{Results for $N_t = 10, 12, 16, 24$}

\begin{figure}[t!]
  \centering
  \includegraphics[width=0.9\linewidth, clip]{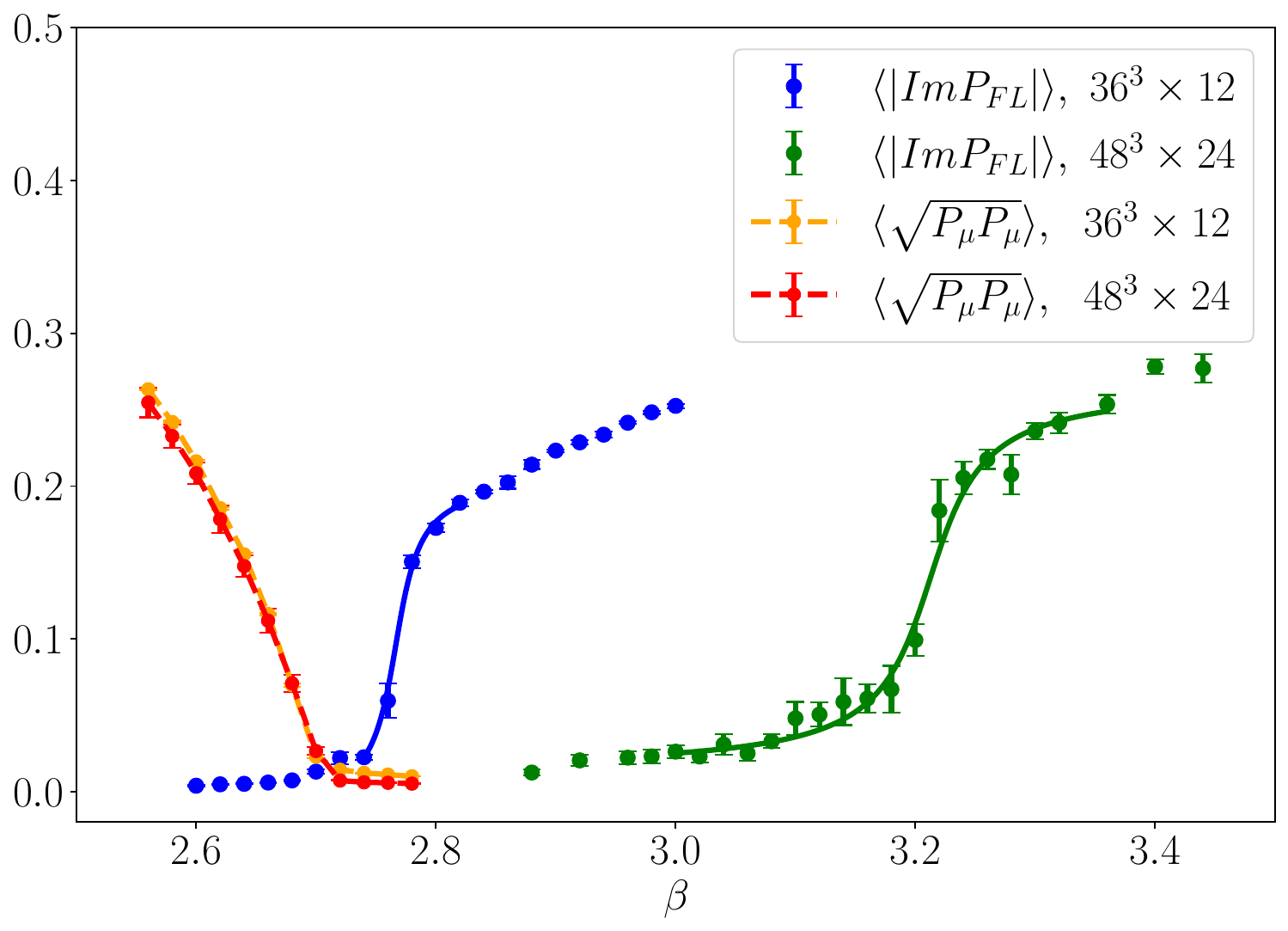}
  \caption{Thermal scaling for the Polyakov loop vs bulk scaling for the exotic phase order parameter. Parameters used: $\hat m=0.0125$, $\hat{\mu} = i\pi$. }
  \label{fig:scaling_polyakov}
\end{figure}

We extend the numerical simulations at imaginary chemical potential to lattices with $N_t = 10, 12, 16, 24$. The behavior of the exotic‑to‑weak‑coupling transition becomes particularly clear at these larger temporal extents, as illustrated in Fig.~\ref{fig:scaling_polyakov}. Although the location of the transition signaled by the Wilson-flowed Polyakov loop shifts when moving from $N_t=12$ (blue) to $N_t=24$ (green), as expected for a thermal transition, the exotic order parameter (orange and red) shows essentially no change, indicating the bulk nature of the associated transition.

\begin{figure}[t!]
  \centering
  \includegraphics[width=0.9\linewidth, clip]{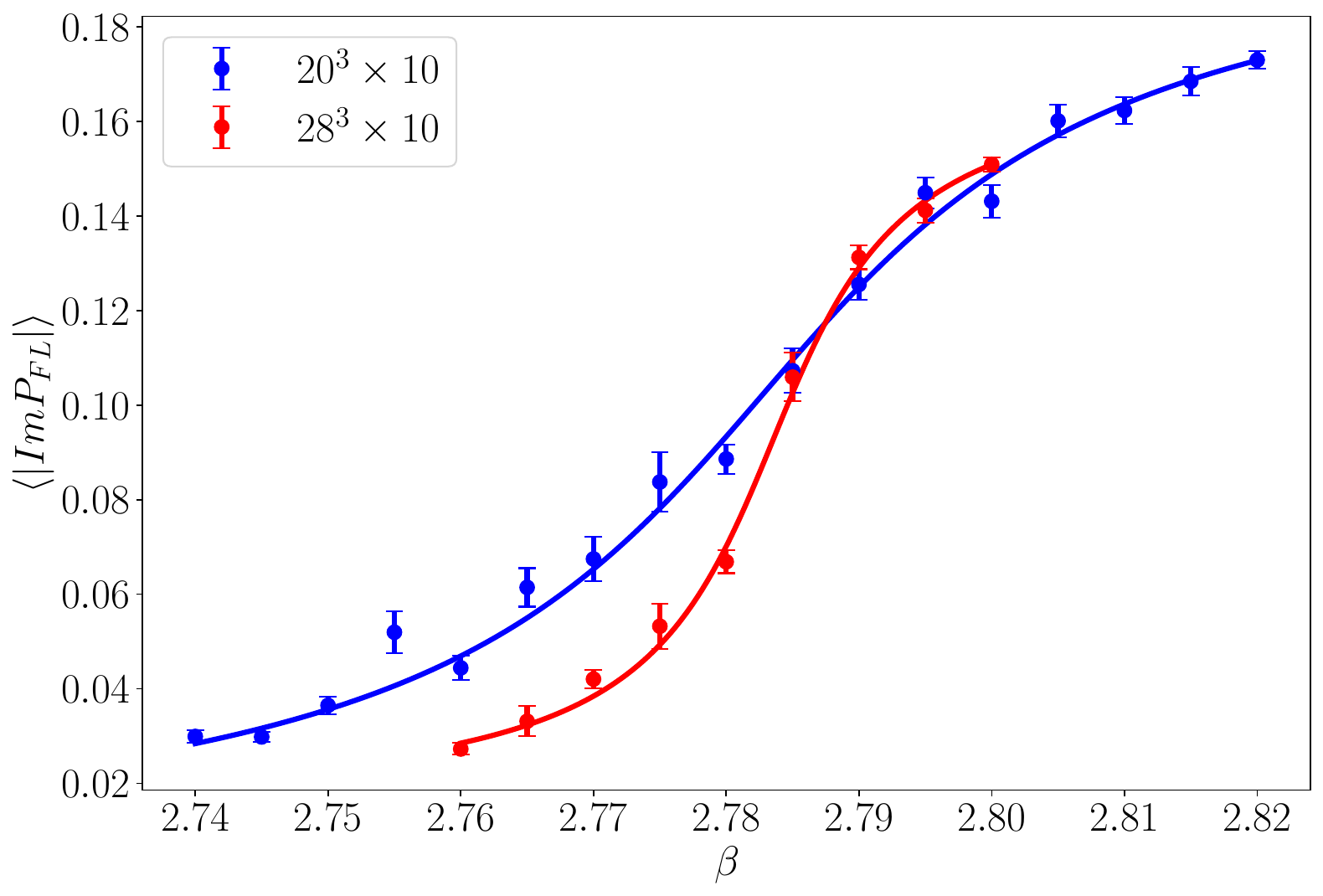}
  \caption{Wilson-flowed Polyakov loop for two different volumes on $N_t=10$ lattices. Parameters used: $\hat m=0.02$, $\hat{\mu} = i \pi.$}
  \label{fig:testvol}
\end{figure}

For $N_t\geq16$, the ratio $N_s/N_t$ was reduced from $3$ to $2$ to save computational time. This could in principle introduce finite-volume effects; however, these effects turn out to be very small. Fig.~\ref{fig:testvol} shows the results of a dedicated test run at $N_t=10$, comparing the imaginary part of the Polyakov loop for $\hat m=0.02$ on two lattices, $20^3 \times 10$ (blue) and $28^3 \times 10$ (red). The resulting critical couplings, $\beta_{RW}=2.786(05)$ and $2.784(05)$, agree within errors, indicating that finite-volume effects are negligible, at least in this case.

\begin{figure}[t!]
  \centering
  \includegraphics[width=0.9\linewidth, clip]{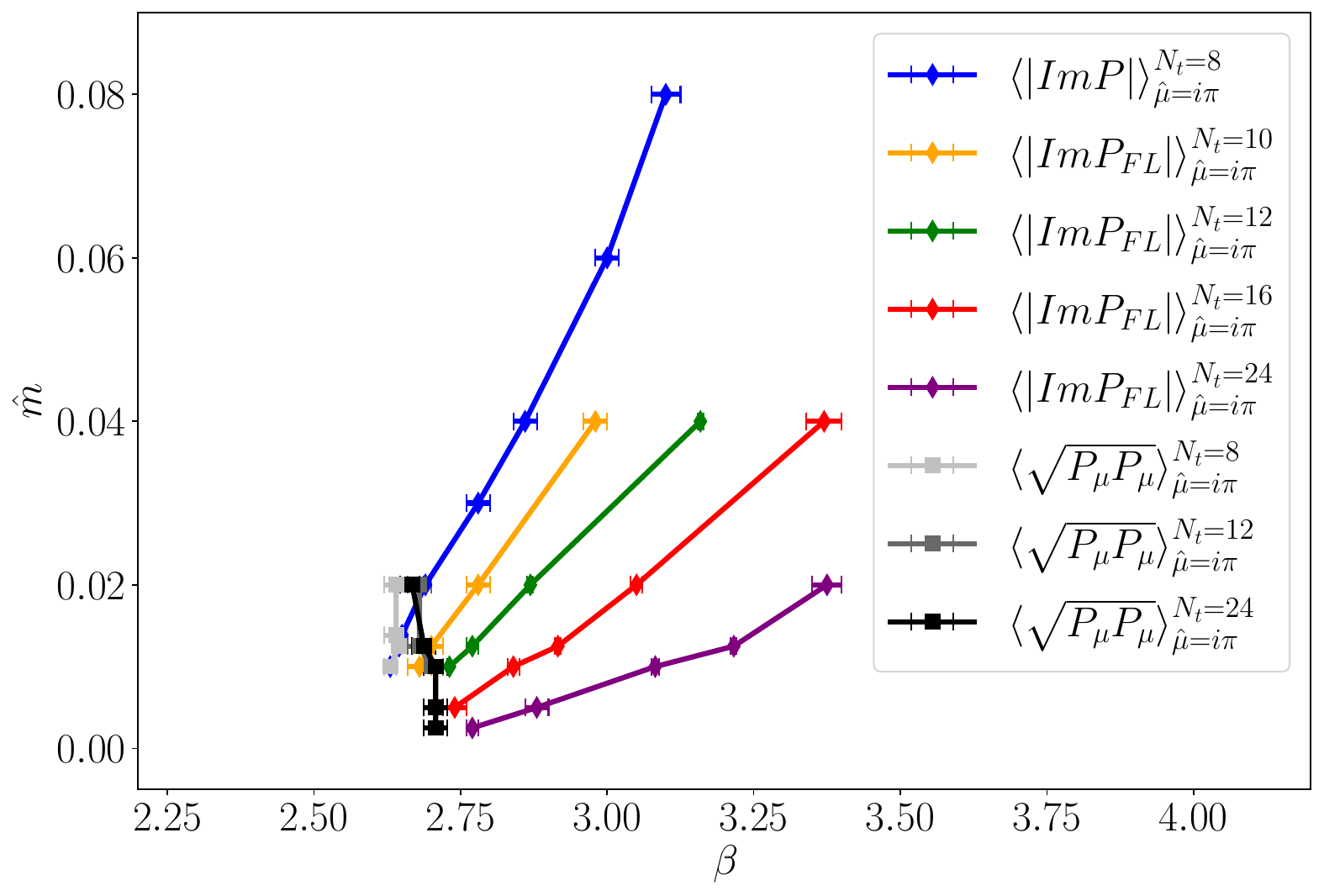}
  \caption{Phase diagram of $N_f=8$ QCD on $N_t=8$--$24$ lattices at $\hat{\mu}= i \pi$. The Roberge-Weiss and bulk transitions are illustrated for different temporal extents respectively by different colors and different shades of gray.}
  \label{fig:cc_l_vs_pp_ntall}
\end{figure}

\begin{figure}[t!]
  \centering
  \includegraphics[width=0.9\linewidth, clip]{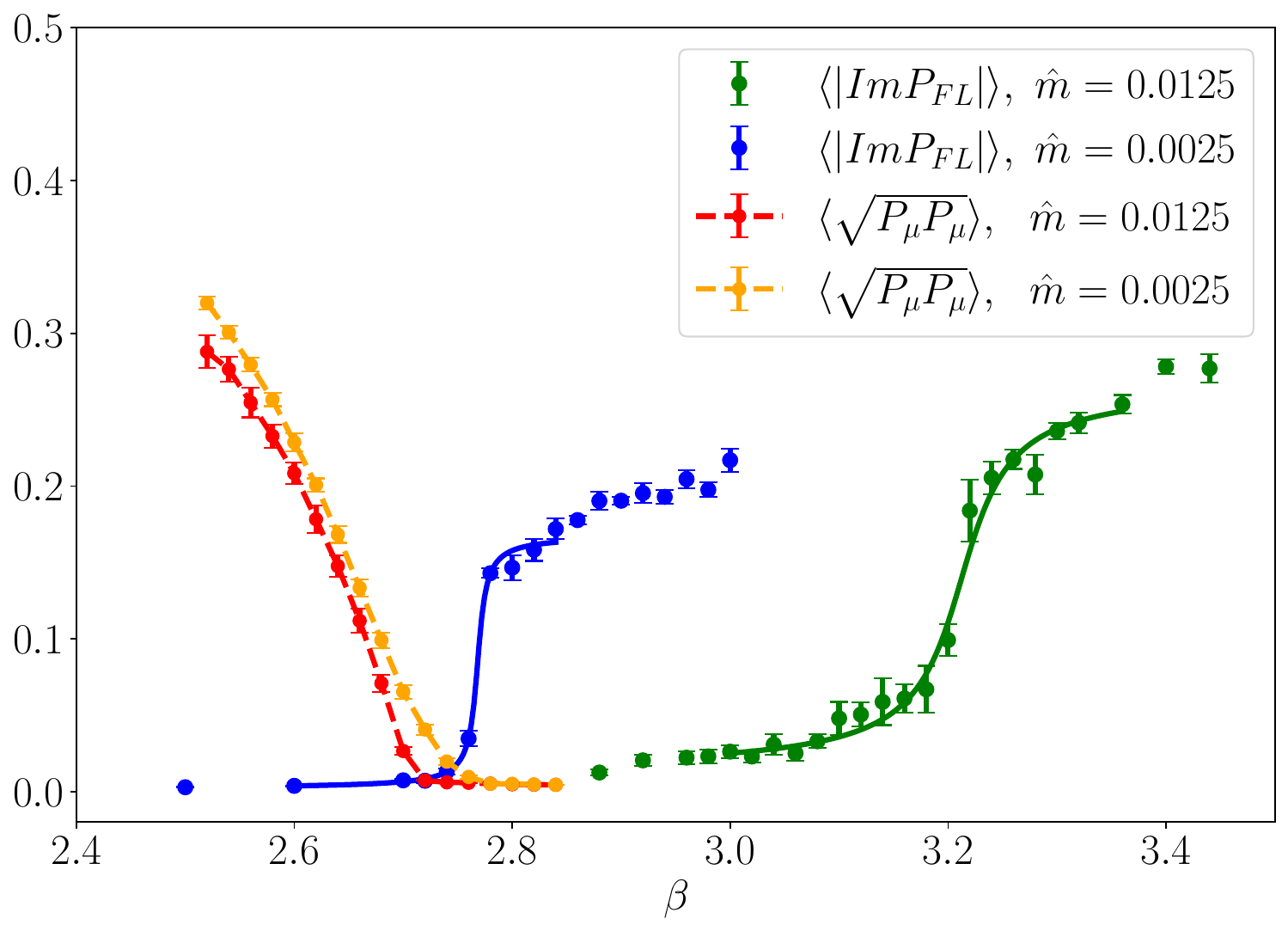}
  \caption{Scaling with the bare quark mass of the RW order parameter vs that of the exotic phase order parameter.
      Data refer to:
    $N_s=48$, $N_t=24$, $\hat{\mu} = i\pi$.}
  \label{fig:scaling_mass}
\end{figure}

Fig.~\ref{fig:cc_l_vs_pp_ntall} summarizes the $(\beta, \hat m)$ phase diagram found for $\hat \mu = i \pi$; data are also reported in Tables~\ref{tab:rw} and \ref{tab:bulk}. The positions of the exotic-to-weak-coupling bulk transition are shown for $N_t=8, 12, 24$ in grey, dark grey and black, respectively. As already noted, the comparison between $N_t=12$ and $24$ highlights the bulk nature of this transition. The Roberge-Weiss transition for $N_t=8, 10, 12, 16, 24$ is shown in blue, orange, green, red and purple, respectively.
One can notice that, at sufficiently small masses, the Roberge-Weiss transition inevitably merges with the bulk transition. As a consequence, the quark mass dependence of $\beta_{RW} (\hat m , N_t)$ becomes increasingly steep as $N_t$ increases, as shown in Fig.~\ref{fig:scaling_mass}.
These results raise once again the question of whether a finite-temperature Roberge-Weiss transition persists at all in the combined chiral and continuum limits.

\begin{table}[ht]
\centering
\begin{tabular}{c | c c c c c}
\hline $\hat m$ & $N_t=8$ & $N_t=10$ & $N_t=12$ & $N_t=16$ & $N_t=24$ \\
\hline $0.002500$ & -- & -- & -- & -- & $2.770(10)$ \\
$0.005000$ & -- & -- & -- & $2.740(20)$ & $2.880(20)$ \\
$0.010000$ & $2.630(10)$ & $2.679(05)$ & $2.731(02)$ & $2.840(10)$ & $3.082(06)$ \\
$0.012500$ & $2.645(03)$ & $2.698(05)$ & $2.770(10)$ & $2.916(05)$ & $3.216(05)$ \\
$0.013844$ & $2.650(10)$ & -- & -- & -- & -- \\
$0.020000$ & $2.690(10)$ & $2.784(05)$ & $2.869(05)$ & $3.050(10)$ & $3.375(25)$ \\ 
$0.030000$ & $2.780(20)$ & -- & -- & -- & -- \\
$0.040000$ & $2.860(20)$ & $2.982(10)$ & $3.159(05)$ & $3.370(30)$ & -- \\
$0.060000$ & $3.000(20)$ & -- & -- & -- & -- \\
$0.080000$ & $3.100(25)$ & -- & -- & -- & -- \\
\hline 
\end{tabular}
\caption{Critical couplings $\beta_{RW}$ of the RW transition.}
\label{tab:rw}
\end{table}

\begin{table}[ht]
\centering
\begin{tabular}{c | c c c c c}
\hline $\hat{m}$ & $N_t=8$ & $N_t=10$ & $N_t=12$ & $N_t=16$ & $N_t=24$ \\
\hline $0.002500$ & -- & -- & -- & -- & $2.700(20)$ \\
$0.005000$ & -- & -- & -- & $2.700(20)$ & $2.700(20)$ \\
$0.010000$ & $2.630(10)$ & $2.677(06)$ & $2.700(20)$ & $2.700(20)$ & $2.700(20)$ \\
$0.012500$ & $2.645(03)$ & $2.675(15)$ & $2.680(20)$ & $2.680(20)$ & $2.680(20)$ \\
$0.013844$ & $2.640(20)$ & -- & -- & -- & -- \\
$0.020000$ & $2.640(20)$ & $2.654(20)$ & $2.680(20)$ & $2.660(20)$ & $2.660(20)$ \\
\hline 
\end{tabular}
\caption{Critical couplings $\beta_B$ of the bulk transition.}
\label{tab:bulk}
\end{table}

\begin{table}[ht]
\centering
\begin{tabular}{c | c }
\hline $N_t$ & $\beta_{RW} (\hat m = 0, N_t)$ \\
\hline
8 &  2.518(44)\\
10 &  2.567(13) \\
12 &  2.593(09)\\
16 &  2.609(29) \\
24 &  2.682(22)\\
\hline 
\end{tabular}
\caption{Critical couplings of the RW transition, extrapolated to the chiral limit.}
\label{tab:rw_chiral}
\end{table}

\section{Discussion and bare quark mass extrapolation}

\label{sec:discussion}

Let us summarize and discuss our results. We have explored the space of the bare lattice parameters, $\beta$ and $\hat m$, for different values of the Euclidean temporal extent $N_t$, determining for each $N_t$ and for various values of $\hat m$ the critical couplings at which the RW and bulk transitions occur, denoted by $\beta_{RW} (\hat m, N_t)$ and $\beta_B (\hat m, N_t)$, respectively.

When $\beta_{RW} > \beta_B$, the RW transition takes place in the physical region of the $(\beta,\hat m)$ plane, which is connected to the continuum theory (by asymptotic freedom) in the $\beta \to \infty$ limit. In this region, the thermal nature of the RW transition is reflected in the clear dependence of $\beta_{RW} (\hat m,N_t)$ on the temporal extent $N_t$, at fixed and finite $\hat m$. This would suggest that $\beta_{RW}$ corresponds to a real physical temperature, $T_{RW} = 1 / (N_t a(\beta_{RW},\hat m))$, with a well-defined continuum limit as $N_t \to \infty$.
By contrast, $\beta_B (\hat m, N_t)$ rapidly approaches, at fixed $\hat m$, an asymptotic value as $N_t \to \infty$, indicating the bulk nature of this transition.

However, as the bare quark mass decreases, $\beta_{RW} (\hat m,N_t)$ rapidly approaches $\beta_B (\hat m, N_t)$ and merges with it for all explored values of $N_t$, before $\hat m = 0$ is reached. This means that, for sufficiently small quark masses, one can decrease $\beta$, and hence the temperature, without ever reaching the low-$T$ confined phase, because the unphysical exotic phase is encountered first. Since the exotic phase is also characterized by an unbroken RW symmetry, this explains why $\beta_{RW}$ and $\beta_B$ coincide.
Still, this does not by itself imply that $T_{RW} = 0$ in the chiral limit:
in principle, one could explore even larger values of $N_t$, to determine whether the situation changes,
but at the price of an excessive computational cost and with the risk of remaining in the same state of uncertainty.

A systematic way to approach the problem would be to work along lines of constant physics, i.e., to locate lines in the physical region of the $(\beta, \hat m)$ plane along which the ratios of hadron masses and other physical quantities remain constant, and then determine the continuum limit of $T_{RW}$ along these lines.
After that, one could study the behavior of the continuum-extrapolated value of $T_{RW}$ as a function of the pion mass, to determine whether any definite conclusion can be reached as the pion mass goes to zero. Such an approach, however, would require a prohibitive computational effort, including costly zero-temperature simulations, and is therefore far beyond the scope of this exploratory investigation.
\\

Given this situation, we explore the following approach. For each fixed value of $N_t$, the behavior of $\beta_{RW} (\hat m,N_t)$ as a function of $\hat m$, before it meets the bulk transition, appears sufficiently smooth to allow for a reliable extrapolation to the $\hat m = 0$ limit. 
One can then regard the resulting extrapolations $\beta_{RW} (\hat m = 0,N_t)$ as estimates of the values that would be obtained in the chiral limit for different values of $N_t$, were it not for the presence of the bulk transition. The key question then is whether their dependence on $N_t$ suggests that $\beta_{RW} (\hat m = 0,N_t)$ eventually exits the bulk region as $N_t \to \infty$, and, if it does, whether this occurs with the $N_t$-scaling expected for a finite $T_{RW}$ in the chiral and continuum limit. 
This approach has, of course, some shortcomings, such as exchanging the chiral and continuum limits \cite{Durr:2004ta,Bernard:2004ab,Kronfeld:2007ek}, or performing the chiral limit at fixed spatial volume. Nevertheless, in light of the considerations above, this appears to be the most viable strategy.

\begin{figure}[ht!]
  \centering
  \includegraphics[width=1.0\linewidth, clip]{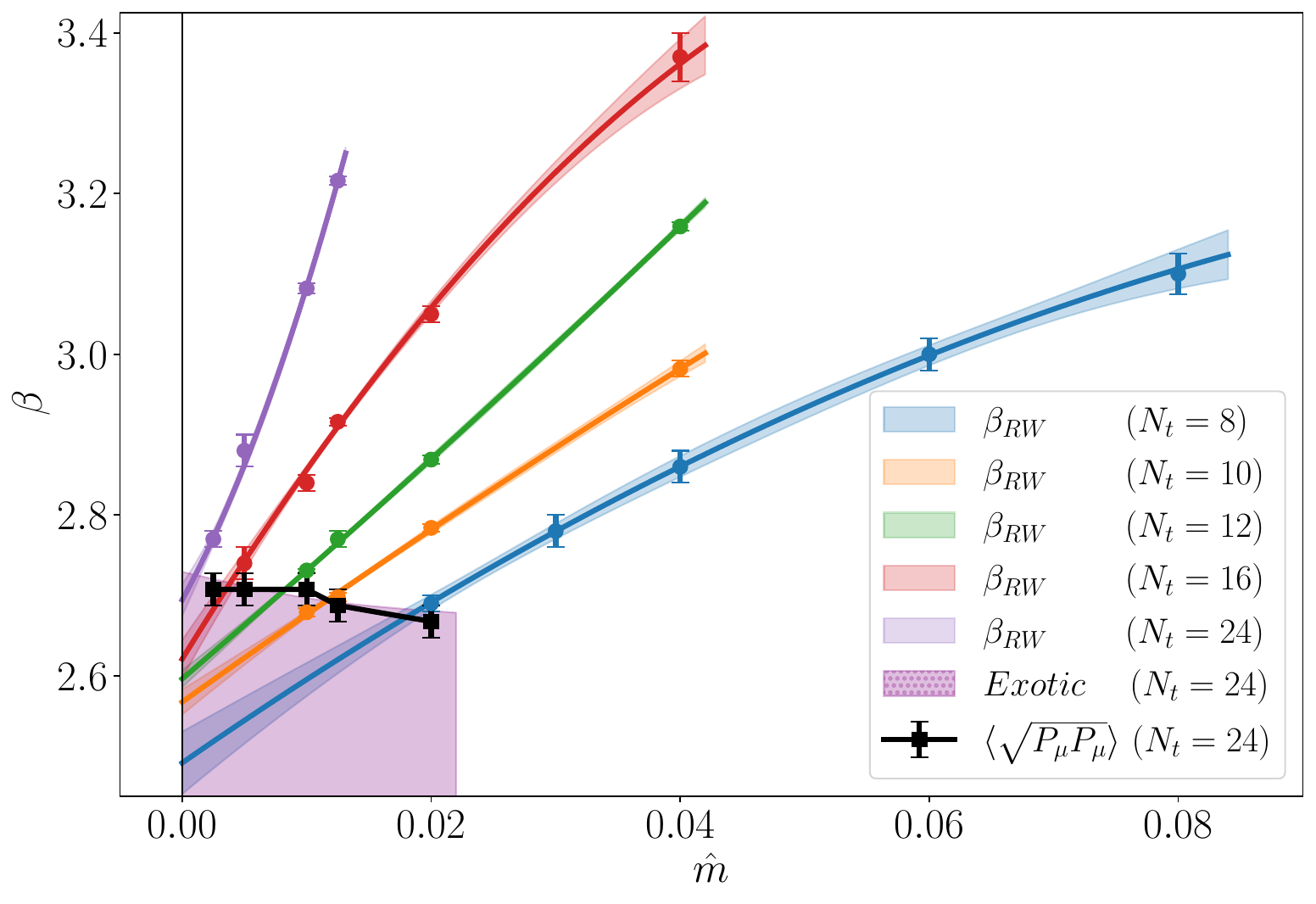}
  \caption{Summary of the $(\hat m, \beta)$ phase diagram, together with extrapolations to the zero bare quark mass limit of the RW lines obtained for each $N_t$ in the physical region of the parameter space. For the bulk transition, only results obtained for $N_t = 24$ are reported. The exotic phase continues also beyond $\hat m = 0.02$; however, the shaded region is truncated there because no further determinations are available at larger masses. }
  \label{fig:phase_diagram}
\end{figure}

\begin{figure}[ht!]
  \centering
  \includegraphics[width=0.9\linewidth, clip]{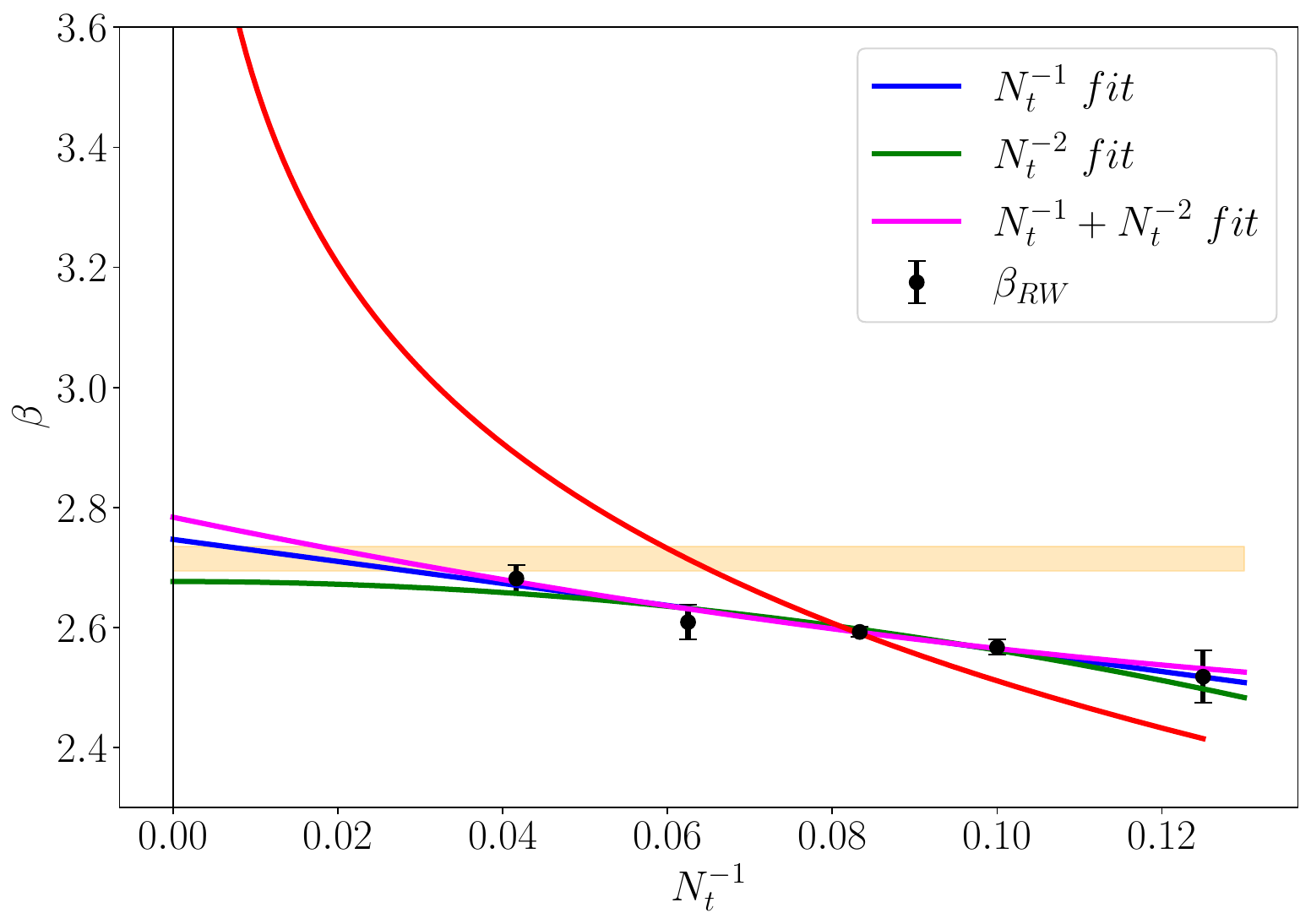}
  \caption{RW couplings in the $\hat m \to 0$ limit as a function of $1/N_t$, compared with the bulk transition (yellow band) and various best fits assuming a finite value in the $N_t \to \infty$ limit, with power law corrections in $1/N_t$. 
  The red line represents instead the behavior that would be expected for a true thermal transition, taking $N_t = 12$ as a reference point and assuming a scaling of the lattice spacing according to the two-loop $\beta$-function.}
  \label{fig:ntfit}
\end{figure}

To proceed along this line, Fig.~\ref{fig:phase_diagram} illustrates the phase diagram with the axes exchanged and includes an extrapolation to the chiral limit, performed separately for each $N_t$. Only the data outside the exotic region are used in the fits, ensuring that the extrapolation is not contaminated by lattice artifact physics.
Note that the figure shows only the boundary of the exotic phase for $N_t=24$ (black), but we have determined this boundary independently for each $N_t$, and in each case the fits include only data points lying outside the corresponding exotic region (however, as already discussed, the determinations for $N_t \geq 12$ agree within errors with the determination for $N_t=24$). 
For the fits we use a linear and a quadratic ans\"atz,
\beq
\beta^{(1)}_{RW}(\hat{m}, N_t)|_{N_t} & = & a^{(1)} + b^{(1)} ~ \hat{m} \nonumber \\
\beta^{(2)}_{RW}(\hat{m}, N_t)|_{N_t} & = & a^{(2)} + b^{(2)} ~ \hat{m} + c^{(2)} ~\hat{m}^2 \mbox{ . }
\label{eq:chiral_extrap_ansatze}
\eeq
The quadratic extrapolations are shown as colored bands in Fig.~\ref{fig:phase_diagram}. The extrapolated values at $\hat m=0$, including a systematic uncertainty estimated from comparing the quadratic and linear fits, are then 
collected in Table~\ref{tab:rw_chiral}, and also shown as a function of 
$1/N_t$ in Fig.~\ref{fig:ntfit}, where we also report
(horizontal band) the boundary of the unphysical exotic phase extrapolated to $\hat m = 0$,
i.e., the chiral extrapolation of the $N_t$-independent bulk transition, $\beta_B = 2.715(25)$.

The extrapolated critical couplings $\beta_{RW} (\hat m = 0,N_t)$ show a clear tendency to remain within 
the unphysical region even as the $N_t \to \infty$ limit is approached.
This is confirmed by best fits assuming a finite value of $\beta_{RW} (\hat m = 0,N_t = \infty)$ and including linear and/or quadratic corrections in $1/N_t$, i.e.,
\beq
\beta_{RW}(\hat{m}=0, N_t) = \beta_{RW}(\hat{m}=0, \infty) + \frac{a}{N_t} + \frac{b}{N_t^2} \, ,
\label{eq:nt_extrap_ansatze}
\eeq
all of which yield acceptable $\chi^2/ndof$ values: $\sim 0.4$ when both corrections are included or when setting $a=0$, and $\sim 0.9$ when setting $b = 0$. These fits are therefore consistent with a finite $N_t \to \infty$ limit, $\beta_{RW}=2.73(5)$, which is compatible, within uncertainties, with lying in the unphysical region\footnote{Analogous results are obtained by fitting corrections to a constant using inverse powers of the total lattice volume.}.

By contrast, for a genuine thermal phase transition, the critical couplings $\beta_{RW}(N_t)$ should scale with $N_t$ in such a way that $T_{RW} = 1 / (N_t a(\beta_{RW}))$ is independent of $N_t$. Equivalently, one should have $N_t^{-1} \propto a(\beta_{RW}(N_t))$, and therefore, by asymptotic freedom, $\beta_{RW}$ should diverge as $N_t \to \infty$. In particular, considering the two-loop $\beta$-function for three colors, one can derive the lattice spacing as a function of the bare coupling, obtaining
\beq
a(\beta) & = & \frac{1}{\, \Lambda_L} R(\beta) \label{eq:scaling_function} \\
R(\beta) & = & \Bigl(\frac{6 b_0}{\beta} \,\Bigr)^{(-b_{1}/2\,b_{0}^2)} \exp\Bigl(-\frac{\beta}{12\,b_{0}}\Bigr), \nonumber
\eeq
where $\Lambda_L$ is the QCD lattice scale parameter, $R(\beta)$ defines the two-loop asymptotic scaling function and $b_0$ and $b_1$ are the universal one- and two-loop coefficients of the $\beta$-function,
\beq
	b_0 & = & \frac{1}{16\pi^2}\left(11 - \frac{2}{3} N_f\right), \nonumber \\
	b_1 & = & \frac{1}{(16\pi^2)^2}\left(102 - \frac{38}{3} N_f\right). \nonumber
\eeq
The expected behavior for a genuine thermal transition~\cite{Deuzeman:2008sc,Deuzeman:2008pf}, $N_t^{-1} = \mathrm{const} \times  R(\beta_{RW}(N_t))$,
is shown in Fig.~\ref{fig:ntfit} as a red line, with the constant chosen so that it matches
the result obtained for $N_t = 12$. It is evident that such behavior is not compatible with our data.

To summarize, the collected evidence is against the existence of a finite temperature RW transition in the chiral limit of the continuum physical theory.
As discussed above, this would in turn imply that QCD with $N_f = 8$ chiral fermions in the fundamental representation already lies in the conformal window, i.e., that $N_f^* \leq 8$.

\FloatBarrier

\section{Conclusions}

\label{sec_conclusions}

In this study, we have considered the problem of identifying the onset of the conformal window in QCD with $N_f$ chiral flavors in the fundamental representation, proposing the investigation of the Roberge-Weiss transition as an effective method to approach such problem by lattice QCD simulations. The idea of studying the thermal transition temperature as a function of $N_f$, in order to find the point where it vanishes in the chiral limit as a manifestation of the onset of conformality, is not new \cite{Cuteri:2017gci,Cuteri:2018wci,Cuteri:2021ikv,Klinger:2025xxb,Klinger:2026pbe}. 
In this respect, the novelty of our proposal lies in the fact that the RW symmetry is exact for any value of the quark mass, so that the determination of $\beta_{RW} (\hat m, N_t)$ and its corresponding chiral extrapolation can be carried out with much greater reliability, since they are based on the analysis of well-defined order parameters. Moreover, we have argued that the critical value of $N_f$ at which $T_{RW}$ vanishes in the chiral limit must coincide with the onset of the conformal window.

As a first implementation of our proposal, we have carried out a numerical investigation of $N_f = 8$ QCD at the RW point, discretized via stout-improved staggered fermions and the tree-level improved Symanzik pure gauge action. In particular, we considered simulations for five different values of the Euclidean temporal extent, $N_t = 8, 10, 12, 16, 24$, determining in each case the critical couplings $\beta_{RW} (\hat m, N_t)$ and $\beta_B (\hat m, N_t)$ associated with the RW transition and the bulk transition, respectively. The latter separates the physical region of the $(\beta,\hat m)$ parameter space, connected to the continuum theory, from the exotic phase where the staggered single-site shift symmetry is spontaneously broken.

We have found that, as long as the critical couplings $\beta_{RW} (\hat m, N_t)$ lie in the physical region, they exhibit the $N_t$-dependence typical of a genuine thermal transition. As the bare quark mass decreases, however, $\beta_{RW} (\hat m,N_t)$ rapidly approaches $\beta_B (\hat m, N_t)$ and merges with it, for each of the explored values $N_t$, before reaching $\hat m = 0$. While this evidence does not support the persistence of a thermal RW transition in the chiral limit, it does not completely exclude it either.

For this reason, to gather further evidence from our results, we considered the $\beta_{RW} (\hat m = 0,N_t)$ extrapolations obtained from the determinations in the physical region. Their dependence on $N_t$, rather than being compatible with a thermal transition, suggests that the $N_t \to \infty$ limit remains finite and within the bulk region.
Therefore, our present results suggest that no thermal RW transition persists in the continuum limit of $N_f = 8$ QCD, i.e., that $T_{RW} = 0$ and the Roberge-Weiss symmetry is spontaneously broken for any temperature. According to our argument, this would mean that $N_f = 8$ is already in the conformal window, a result supported by some previous evidence reported in the literature \cite{NunesdaSilva:2015jpf,Hasenfratz:2022zsa,Witzel:2024bly}.

Our present numerical results should be regarded as exploratory, since they may be affected by various sources of systematic uncertainty; in particular, they are based on chiral extrapolations performed at finite lattice spacing and, owing to computational limitations, on lattices of only moderate size, with aspect ratios limited to 2-3.
Moreover, the advantage of dealing with a real phase transition for all values of the quark mass has not been fully 
exploited, since we have not performed a finite size scaling (FSS) analysis to achieve a precise
determination of the critical couplings in the thermodynamical limit.
In this respect, future investigations should consider larger volumes, include a FSS analysis, and possibly perform continuum extrapolations along lines of constant physics.

Nevertheless, our proposal provides a new and effective approach to determining 
the extent of the conformal window in QCD with quarks in the fundamental representation,
which should be refined in the future by extending our investigation to different values
  of $N_f$, in particular to those around $N_f = 8$.

\acknowledgements

We thank C.~Bonanno, M.P.~Lombardo, A.~Hasenfratz and A.~Vichi for useful discussions.
This work has been partially supported
by the project “Non-perturbative aspects of fundamental interactions, in the Standard Model and beyond” funded by MUR,
Progetti di Ricerca di Rilevante Interesse Nazionale (PRIN), Bando 2022, grant 2022TJFCYB (CUP I53D23001440006).
Numerical simulations have been performed on Leonardo at CINECA, based on the agreement between INFN and CINECA under projects INF24\_npqcd and INF25\_npqcd. KZ acknowledges support by the
project “Non-perturbative aspects of fundamental interactions, in the Standard Model and beyond” funded by MUR,
Progetti di Ricerca di Rilevante Interesse Nazionale (PRIN), Bando 2022, Grant 2022TJFCYB (CUP I53D23001440006).

\bibliography{biblio}

\end{document}